\documentclass[conference,english]{IEEEtran}

\usepackage[T1]{fontenc}
\usepackage[latin9]{inputenc}

\usepackage{mathrsfs}

\usepackage{cite}

\usepackage{amsthm}
\usepackage{mathtools} 
\usepackage{amssymb}

\usepackage{graphicx}

\usepackage{babel}

\usepackage{flushend}

\makeatletter

\theoremstyle{plain}
\newtheorem{theorem}{\protect\theoremname}

\theoremstyle{plain}
\newtheorem{lemma}[theorem]{\protect\lemmaname}

\theoremstyle{plain}
\newtheorem{corollary}[theorem]{\protect\corollaryname}
\theoremstyle{plain}

\makeatother

\providecommand{\corollaryname}{Corollary}
\providecommand{\lemmaname}{Lemma}
\providecommand{\theoremname}{Theorem}
\providecommand{\propositionname}{Proposition}

\renewcommand{\IEEEQED}{\IEEEQEDopen} 	

\DeclareMathOperator{\pr}{\mathbf{P}} 		

\begin{document}

\title{
  On the Cost and Benefit of Cooperation\\
  (Extended Version)     
  }

\author{\IEEEauthorblockN{Parham Noorzad}
\IEEEauthorblockA{
California Institute of Technology\\
parham@caltech.edu}
\and
\IEEEauthorblockN{Michelle Effros}
\IEEEauthorblockA{
California Institute of Technology\\
effros@caltech.edu}
\and
\IEEEauthorblockN{Michael Langberg}
\IEEEauthorblockA{
State University of New York at Buffalo\\
mikel@buffalo.edu}}

\maketitle

\begin{abstract}
In a cooperative coding scheme, network nodes work together to
achieve higher transmission rates. 
To obtain a better understanding of cooperation,
we consider a model in which two transmitters send 
rate-limited descriptions of their messages to a ``cooperation facilitator'',
a node that
sends back rate-limited descriptions of the pair to each transmitter.
This model includes the conferencing encoders model 
and a prior model from the current authors as special cases. We show that
except for a special class of multiple access channels, the gain in sum-capacity
resulting from cooperation under this model is quite large. Adding a
cooperation facilitator to any such channel results in a network that does
not satisfy the edge removal property. An important special case is the
Gaussian multiple access channel, for which we explicitly characterize the
sum-rate cooperation gain.
\end{abstract}

\section{Introduction} \label{sec:intro}

To meet the growing demand for higher transmission rates, network nodes should 
employ coding schemes that use scarce resources
in a more efficient manner. By working together,
network nodes can take advantage of under-utilized network resources to help data 
transmisssion in heavily constrained regions of the network.
Cooperation among nodes emerges as a natural 
strategy towards this aim.

We propose a network model and use it to study the cost and benefit
of enabling a cooperation in a given network.
As an example, consider two nodes, $A$ and $B$, transmitting 
independent messages over a network $\mathcal{N}$. 
A third node $C$ that has bidirectional links to $A$ and $B$ can help $A$
and $B$ work together to achieve a higher sum-rate than they would have
achieved had they worked separately. 

\begin{figure} 
	\begin{center}
		\includegraphics[scale=0.25]{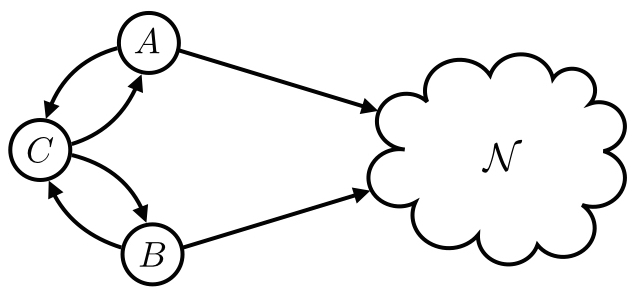}
		\caption{An example of cooperation among network nodes.
		Node $C$ enables nodes $A$ and $B$ to cooperate and potentially 
		achieve higher rates in the transmission of their messages over 
		network $\mathcal{N}$.} \label{fig:example}
	\end{center}
\end{figure}

We seek to understand how the gain in sum-rate resulting
from cooperation between $A$ and $B$  relates to the capacities of the links
from $(A,B)$ to $C$ and back. Intuitively, we think of the increase in sum-rate
as the \emph{benefit} of cooperation and the capacities of the links between $(A,B)$
and $C$ as the \emph{cost} of cooperation. See Figure \ref{fig:example}.

\begin{figure} 
	\begin{center}
		\includegraphics[scale=0.19]{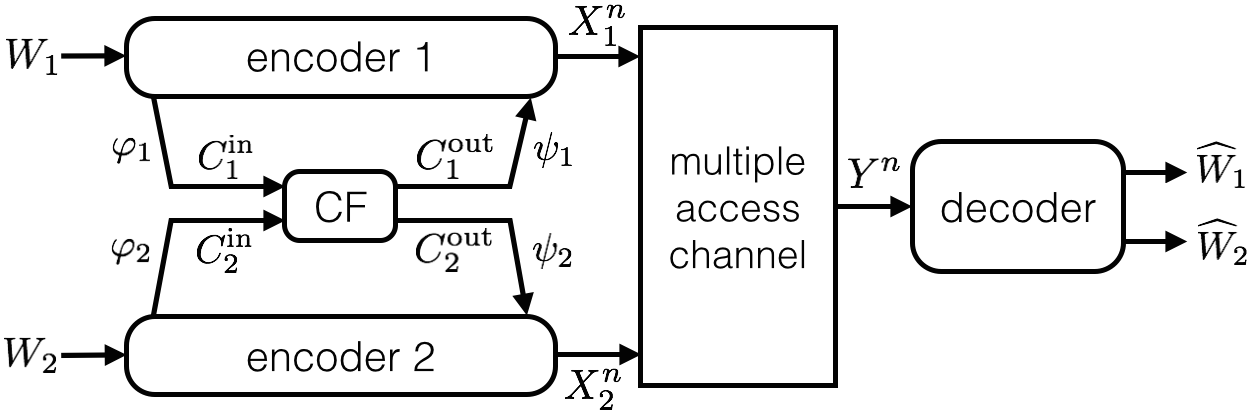}
		\caption{The network model for the MAC with a CF.}\label{fig:networkmodel}
	\end{center}
\end{figure}

To study this situation formally, let $A$ and $B$ be the encoders
of a memoryless multiple access channel (MAC).
Let $C$ be a ``cooperation facilitator'' (CF), a node which,
prior to the transmission of the messages over the network, receives a 
rate-limited description of each encoder's message and sends a rate-limited 
output to each encoder. See Figure \ref{fig:networkmodel}.

In \emph{one-step cooperation}, each encoder sends a function of its message 
to the CF and the CF transmits, to each encoder, a value that is a function 
of both of its inputs. Similarly, we can define 
$k$-\emph{step cooperation} (for a fixed positive integer $k$) between the 
CF and the encoders where the information transmission between the CF and each 
encoder continues for $k$ steps, with the constraint that the
information that the CF or each encoder transmits in each step only depends on the information
that it previously received. In our achievability result, however, we only use
one-step cooperation for simplicity.

Our CF extends the cooperation model introduced by a previous work 
of the authors \cite{NoorzadEtAl} to allow for rate-limited inputs. 
While the CF in \cite{NoorzadEtAl} has full knowledge of both 
messages and transmits a rate-limited output to both encoders, 
the more general CF we study in this paper only has partial knowledge of each 
encoder's message. In addition, unlike in \cite{NoorzadEtAl}, 
we allow the CF to send a different output to each encoder. We define
our cooperation model formally in Section \ref{sec:coop}.

The main result of \cite{NoorzadEtAl} states that there exists a
discrete memoryless MAC where encoder cooperation through a CF
results in a large gain (with respect to the capacities of the
output edges of the CF). This implies the existence of a 
network consisting of a MAC with a CF that does not satisfy the
``edge removal property'' \cite{HoEtAl,JalaliEtAl}. 
We say a network satisfies the edge removal property if
removing an edge from that network does not reduce the achievable rate of any
of the source messages by more than the capacity of that edge. A 
question that remained unanswered in \cite{NoorzadEtAl} was whether such
a result is true for more natural channels, e.g., the Gaussian MAC.
The answer turns 
out to be positive, and except for a special class of MACs,
adding a CF results in a large sum-capacity gain (Theorem \ref{thm:growthrate}).

Our achievability scheme
combines three coding schemes via rate splitting.
First, each encoder sends part of its message to the CF. 
The CF passes on part of what it receives from each
encoder to the other encoder without any further operations. In this
way the CF enables ``conferencing'' between the encoders, which is a
cooperation strategy introduced by Willems \cite{Willems}. 

The CF uses the remaining part of what it receives to help the 
encoders ``coordinate'' their transmissions;
that is, it enables the encoders to create dependence among 
independently generated codewords.
For this coordination strategy, we rely on results from rate-distortion 
theory \cite[pp. 318-324]{CoverThomas}, which were used by Marton \cite{Marton}
and El Gamal and Van der Meulen \cite{ElGamalMeulen} to obtain
an inner bound for the capacity region of the broadcast 
channel. 

Finally, for the remaining part of the messages, which do
not go through the CF, the encoders use the classical coding scheme of 
Ahlswede \cite{Ahlswede1,Ahlswede2} and Liao \cite{Liao}. 
We formally introduce our achievable scheme in Section \ref{sec:ach}, and study
its performance in Section \ref{sec:error}.

In Section \ref{sec:gaussian} we provide an inner bound for the Gaussian MAC with
transmitter cooperation using methods similar to \cite{Wigger}. We compare the
sum-rate gain of our inner bound with the sum-rate gain of schemes 
that use only one or another of our strategies. We see that none alone performs as 
well as their combination, which is the scheme we propose here. 

\section{The Cooperation Model} \label{sec:coop}

Let $(\mathcal{X}_1\times\mathcal{X}_2,P(y|x_1,x_2),\mathcal{Y})$ denote a
memoryless MAC. Suppose $W_1$ and $W_2$ are the messages that 
encoders 1 and 2 transmit, respectively. For every positive integer $k$,
define $[k] =\{1,\dots,k\}$. We assume that $W_1$ and $W_2$ are
independent and uniformly distributed over the sets $[M_1]$ and $[M_2]$, 
respectively. 

For $i=1,2$, represent encoder $i$ by the mappings
\begin{align*}
  \varphi_i &: [M_i] \rightarrow\big[2^{nC_i^\text{in}}\big]\\
  f_i    &: [M_i]\times \big[2^{nC_i^\text{out}}\big] \rightarrow \mathcal{X}_i^n
\end{align*}
that describe the transmissions to the CF and channel, respectively. 
We represent the CF by the mappings
\begin{equation*}
  \psi_i:\big[2^{nC_1^\text{in}}\big]\times\big[2^{nC_2^\text{in}}\big] 
  \rightarrow\big[2^{nC_i^\text{out}}\big],
\end{equation*}
where $\psi_i$ denotes the output of the CF to encoder $i$ for $i=1,2$. 
Under this definition, when $(W_1,W_2)=(w_1,w_2)$, 
the CF receives $\varphi_1(w_1)$ and $\varphi_2(w_2)$ from encoders 1 and 2, 
respectively. The CF then sends $\psi_1(\varphi_1(w_1),\varphi_2(w_2))$
to encoder 1 and $\psi_2(\varphi_1(w_1),\varphi_2(w_2))$ to encoder 2.

We represent the decoder by the mapping
\begin{equation*}
g:\mathcal{Y}^n\rightarrow [M_1]\times[M_2].
\end{equation*}
Then the probability of error is given by
\begin{equation*}
  P_e^{(n)}=\pr\big\{g(Y^n)\neq (W_1,W_2)\big\}.
\end{equation*}

Define $\mathbf{C}^\mathrm{in} =(C_1^\mathrm{in},C_2^\mathrm{in})$
and $\mathbf{C}^\mathrm{out} =(C_1^\mathrm{out},C_2^\mathrm{out})$.
We call the mappings $(\varphi_1,\varphi_2,\psi_1,\psi_2,f_1,f_2,g) $
an $(n,M_1,M_2)$ code for the MAC with a 
$(\mathbf{C}^\mathrm{in},\mathbf{C}^\mathrm{out})$-CF. 
For nonnegative real numbers $R_1$ and $R_2$, we say that the rate pair $(R_1,R_2)$ 
is achievable if for every $\epsilon>0$ and sufficiently large $n$,
there exists an $(n,M_1,M_2)$ code such that $P_e^{(n)}\leq\epsilon$ 
and
\begin{equation*}
  \frac{1}{n}\log M_{i}>R_i-\epsilon,
\end{equation*}
for $i=1,2$. We define the capacity region as the closure of the set
of all achievable rate pairs $(R_1,R_2)$ and denote it by 
$\mathscr{C}(\mathbf{C}^\mathrm{in},\mathbf{C}^\mathrm{out})$.

Using the capacity region of the MAC with conferencing encoders 
\cite{Willems} (Appendix \ref{app:conf}), 
we obtain inner and outer bounds for the capacity
region of a MAC with a CF. Let $\mathscr{C}_\mathrm{conf}(C_{12},C_{21})$
denote the capacity region of a MAC with a $(C_{12},C_{21})$ conference. 
Since the conferencing capacity region can be achieved with a single 
step of conferencing \cite{Willems}, it follows that
\begin{equation*}
  \mathscr{C}_\mathrm{conf}\big(\min\{C_1^\mathrm{in},C_2^\mathrm{out}\},
  \min\{C_2^\mathrm{in},C_1^\mathrm{out}\}\big)
\end{equation*}
is an inner bound for 
$\mathscr{C}(\mathbf{C}^\mathrm{in},\mathbf{C}^\mathrm{out})$. 
In addition, since each encoder could calculate the CF output
if it only knew what the CF received from the other encoder, 
$\mathscr{C}_\mathrm{conf}(C_1^\mathrm{in},C_2^\mathrm{in})$
is an outer bound for 
$\mathscr{C}(\mathbf{C}^\mathrm{in},\mathbf{C}^\mathrm{out})$. We henceforth refer to 
these inner and outer bounds as the \emph{conferencing bounds}.
Note that when $C_2^\text{out}\geq C_1^\text{in}$ and 
$C_1^\text{out}\geq C_2^\text{in}$, the conferencing inner and outer bounds agree,
giving
\begin{equation*}
  \mathscr{C}(\mathbf{C}^\mathrm{in},\mathbf{C}^\mathrm{out})
	=\mathscr{C}_\mathrm{conf}(C_1^\mathrm{in},C_2^\mathrm{in}).
\end{equation*}

We next discuss the main result of this paper. 
For any memoryless MAC 
$(\mathcal{X}_1\times \mathcal{X}_2,P(y|x_1,x_2),\mathcal{Y})$ with 
a $(\mathbf{C}^\mathrm{in},\mathbf{C}^\mathrm{out})$-CF, define
the sum-capacity as
\begin{equation*}
  C_\mathrm{sum}=\max_{\mathscr{C}(\mathbf{C}_\mathrm{in},\mathbf{C}_\mathrm{out})}
	(R_1+R_2).
\end{equation*}
For a fixed $\mathbf{C}_\mathrm{in}$ with
$\min\{C_1^\mathrm{in},C_2^\mathrm{in}\}>0$, define the
``sum-capacity gain''  $G:\mathbb{R}_{\geq 0}\rightarrow \mathbb{R}_{\geq 0}$
as 
\begin{equation*}
  G(C_\mathrm{out})=C_\mathrm{sum}
	(\mathbf{C}^\mathrm{in},\mathbf{C}^\mathrm{out})
	-C_\mathrm{sum}
	(\mathbf{C}^\mathrm{in},\mathbf{0}),
\end{equation*}
where $\mathbf{C}_\mathrm{out}=(C_\mathrm{out},C_\mathrm{out})$
and $\mathbf{0}=(0,0)$. Note that when $C_\mathrm{out}=0$, no cooperation
is possible, thus 
\begin{equation*}
  C_\mathrm{sum}
	(\mathbf{C}^\mathrm{in},\mathbf{0})=
	\max_{P(x_1)P(x_2)}I(X_1,X_2;Y).
\end{equation*}

The next theorem states that for any MAC where using dependent codewords 
(instead of independent ones) results in an increase in sum-capacity, 
the effect of cooperation through a CF can be quite large. In particular, it shows that the network
consisting of any such MAC and a CF does not satisfy the edge removal
property \cite{HoEtAl,JalaliEtAl}.  

\begin{theorem}[Sum-capacity] \label{thm:growthrate}
For any discrete memoryless MAC 
$(\mathcal{X}_1\times \mathcal{X}_2,P(y|x_1,x_2),\mathcal{Y})$
that satisfies
\begin{equation*}
  \max_{P(x_1,x_2)}I(X_1,X_2;Y)
	> \max_{P(x_1)P(x_2)}I(X_1,X_2;Y),
\end{equation*}
we have $G'(0)=\infty$. For the Gaussian MAC, 
a stronger result holds: For some positive 
constant $\alpha$ and sufficiently small
$C_\mathrm{out}$, 
\begin{equation*}
  G(C_\mathrm{out})\geq \alpha\sqrt{C_\mathrm{out}},
\end{equation*}

\end{theorem}

The proof of Theorem \ref{thm:growthrate} (Appendix \ref{app:growthrate}), 
is based on our achievability result for the MAC with
a CF, which we next describe. 
Define 
\begin{equation*}
  \mathscr{R}(\mathbf{C}^\mathrm{in},\mathbf{C}^\mathrm{out})
\end{equation*}
as the set of all rate pairs $(R_1,R_2)$ that for 
$(i,j)\in\{(1,2),(2,1)\}$ satisfy
\begin{align*}
  R_i &< I(X_i;Y|U,V_1,V_2,X_j)+C_i^\mathrm{in}\\
	R_i &< I(X_i;Y|U,V_j,X_j)+C_{i0} \\
	R_1+R_2 &< I(X_1,X_2;Y|U,V_1,V_2)
	+C_1^\mathrm{in}+C_2^\mathrm{in} \\
	R_1+R_2 &< I(X_1,X_2;Y|U,V_i)+C_i^\mathrm{in}+C_{j0}\\
	R_1+R_2 &< I(X_1,X_2;Y|U)+C_{10}+C_{20}\\
	R_1+R_2 &< I(X_1,X_2;Y), 
\end{align*}
for nonnegative constants $C_{10}$ and $C_{20}$, and distributions 
$P(u,v_1,v_2)P(x_1|u,v_1)P(x_2|u,v_2)$ that satisfy
\begin{align}
  C_{i0} &\leq \min\big\{C_i^\mathrm{in},C_j^\mathrm{out}\big\} \label{eq:ci0}\\
	I(V_1;V_2|U)&\leq(C_1^\mathrm{out}-C_{20})+(C_2^\mathrm{out}-C_{10}).\nonumber
\end{align}

In the above definition, the pair $(U,V_i)$ represents the information
encoder $i$ receives from the CF. In addition, the pair $(C_{10},C_{20})$ 
indicates the amount of rate being used on the CF links to enable the
conferencing strategy. The remaining part of rate on the CF links is used to
create dependence between $V_1$ and $V_2$.
 
\begin{theorem}[Achievability] \label{thm:main} For any memoryless MAC 
$(\mathcal{X}_1\times\mathcal{X}_2,P(y|x_1,x_2),\mathcal{Y})$
with a 
$(\mathbf{C}^\mathrm{in},\mathbf{C}^\mathrm{out})$-CF, the rate region
$\mathscr{R}(\mathbf{C}^\mathrm{in},\mathbf{C}^\mathrm{out})$ 
is achievable.
\end{theorem}

A nontrivial special case is the case where the CF has complete
knowledge of both source messages, that is,
$C_1^\mathrm{in}=C_2^\mathrm{in}=\infty$. 
In this case, it is not hard to see (Appendix \ref{app:infty}) that
$\mathscr{R}(\mathbf{C}^\mathrm{in},\mathbf{C}^\mathrm{out})$
simplifies to the set of all nonnegative rate pairs $(R_1,R_2)$
that satisfy
\begin{align*}
	R_1 &< I(X_1;Y|U,X_2)+C_{10} \\
	R_2 &< I(X_2;Y|U,X_1)+C_{20} \\
	R_1+R_2 &< I(X_1,X_2;Y|U)+C_{10}+C_{20}\\
	R_1+R_2 &< I(X_1,X_2;Y), 
\end{align*}
for nonnegative constants $C_{10} \leq C_2^\mathrm{out}$
and $C_{20} \leq C_1^\mathrm{out}$,
and distributions $P(u,x_1,x_2)$ with
\begin{equation*}
  I(X_1;X_2|U)\leq(C_1^\mathrm{out}-C_{20})+(C_2^\mathrm{out}-C_{10}).
\end{equation*}

Note that in this case,
increasing the number of cooperation steps (Section \ref{sec:intro}) 
does not change the family of functions
the CF can compute. Thus as with the case where
$C_1^\mathrm{in}\leq C_2^\mathrm{out}$ and $C_2^\mathrm{in}\leq C_1^\mathrm{out}$,
using more than one step
for cooperation does not enlarge the capacity region. 

The rate region, 
$\mathscr{R}(\mathbf{C}^\mathrm{in},\mathbf{C}^\mathrm{out})$,
in addition to being achievable, is also convex. To prove this,
we show a slightly stronger result. For every $\lambda\in (0,1)$,
$(\mathbf{C}_a^\mathrm{in},\mathbf{C}_a^\mathrm{out})$, and
$(\mathbf{C}_b^\mathrm{in},\mathbf{C}_b^\mathrm{out})$, define
\begin{equation*}
  \mathscr{R}_\lambda =
	\mathscr{R}\big(
	\lambda\mathbf{C}_a^\mathrm{in}+(1-\lambda)\mathbf{C}_b^\mathrm{in},
	\lambda\mathbf{C}_a^\mathrm{out}+(1-\lambda)\mathbf{C}_b^\mathrm{out}
	\big).
\end{equation*}
Also define 
$\mathscr{R}_a=\mathscr{R}
(\mathbf{C}_a^\mathrm{in},\mathbf{C}_a^\mathrm{out})$
and
$\mathscr{R}_b=\mathscr{R}
(\mathbf{C}_b^\mathrm{in},\mathbf{C}_b^\mathrm{out})$. We then have the 
following result.
\begin{theorem}[Convexity] \label{thm:convexity} For any $\lambda\in (0,1)$,
\begin{equation*}
  \mathscr{R}_\lambda \supseteq \lambda\mathscr{R}_a+(1-\lambda)\mathscr{R}_b.
\end{equation*}
\end{theorem}

The addition in Theorem \ref{thm:convexity} is the Minkowski sum 
\cite{Schneider}, defined for any two subsets $A$ and $B$ of $\mathbb{R}^2$ as 
\begin{equation*}
  A+B =\big\{(a_1+b_1,a_2+b_2)|(a_1,a_2)\in A, (b_1,b_2)\in B\big\}.
\end{equation*}

If we set $\mathbf{C}_a^\mathrm{in}=\mathbf{C}_b^\mathrm{in}$
and $\mathbf{C}_a^\mathrm{out}=\mathbf{C}_b^\mathrm{out}$ in Theorem \ref{thm:convexity}
we get $\mathscr{R}\supseteq \lambda \mathscr{R}
+(1-\lambda)\mathscr{R}$, which is equivalent to the convexity 
of $\mathscr{R}$. Using a time-sharing 
argument, we see that the capacity region 
$\mathscr{C}(\mathbf{C}^\mathrm{in},\mathbf{C}^\mathrm{out})$
also satisfies the property stated in Theorem \ref{thm:convexity}.
We prove Theorem \ref{thm:convexity} in Appendix \ref{app:convexity} 
using techniques from the work of Cover, El Gamal, and Salehi \cite{CoverElGamal}.

\section{The Achievability Scheme} \label{sec:ach}

In this section, we give a formal description of our
coding scheme. 
First, pick nonnegative constants $C_{10}$ and $C_{20}$ such that
Equation (\ref{eq:ci0}) holds for $\{i,j\}=\{1,2\}$.
In our achievability scheme, the first $nC_{i0}$ bits of $W_i$
are sent directly from encoder $i$ to encoder $j$
through the CF without any modification. 
We thus require $C_{i0}$ to satisfy inequality (\ref{eq:ci0}). 

Next, choose $C_{1d}$ and $C_{2d}$ such that
\begin{equation} \label{eq:cd}
\begin{aligned}
  C_{1d} &\leq C_1^\mathrm{out}-C_{20}\\
	C_{2d} &\leq C_2^\mathrm{out}-C_{10}.
\end{aligned}
\end{equation}
The values of $C_{1d}$ and $C_{2d}$ specify the amount
of rate used on each of the output links for the coordination
strategy. Finally, choose an input distribution
$P(u,v_1,v_2)P(x_1|u,v_1)P(x_2|u,v_2)$ so that $P(u,v_1,v_2)$
satisfies
\begin{equation} \label{eq:zeta}
  \zeta:= C_{1d}+C_{2d}-I(V_1;V_2|U)>0.
\end{equation}

Fix $\epsilon>0$. Let $A_\epsilon^{(n)}$ be the weakly typical set
\cite[p. 521]{CoverThomas} with respect to the distribution
\begin{equation*}
  P(u,v_1,v_2)P(x_1|u,v_1)P(x_2|u,v_2)P(y|x_1,x_2).
\end{equation*}
By Cram\'{e}r's large deviation theorem \cite[p. 27]{DemboZeitouni}, 
there exists a nondecreasing function 
$\Theta:\mathbb{R}^{+}\rightarrow\mathbb{R}^{+}$
such that
\begin{equation} \label{eq:cramer}
  \pr\Big\{\big(A_\epsilon^{(n)}\big)^c\Big\}
	\leq 2^{-n\Theta(\epsilon)}.
\end{equation}
Fix $\delta>0$ 
and let $A_\delta^{(n)}$ denote the weakly typical set with respect to
$P(u,v_1,v_2)$. We make use of the typical sets $A_\delta^{(n)}$
and $A_\epsilon^{(n)}$ in the encoding and decoding processes, 
respectively. 

We next describe the codebook generation. 
For $i=1,2$, let
$M_{i}=\lfloor2^{nR_i}\rfloor$ and define
$R_{i0}= \min\{R_i,C_{i0}\}$, $R_{id}=\min\{R_i,C_i^\mathrm{in}\}-R_{i0}$,
and $R_{ii}=(R_i-C_i^\mathrm{in})^+$, where for any 
real number $x$, $x^+=\max\{x,0\}$. Note that for 
$i=1,2$, $R_i=R_{i0}+R_{id}+R_{ii}$, thus we can
split each of the messages into three parts as 
\begin{equation*}
  W_i = (W_{i0},W_{id},W_{ii})
  \in \big[2^{nR_{i0}}\big]\times\big[2^{nR_{id}}\big]\times
  \big[2^{nR_{ii}}\big].
\end{equation*}
Here $W_{10}$ and $W_{20}$ are used for conferencing, 
$W_{1d}$ and $W_{2d}$ are used for coordination, and
$W_{11}$ and $W_{22}$ are transmitted over the channel independently.

Next, for every $(w_{10},w_{20})\in [2^{n(R_{10}+R_{20})}]$, 
generate $U^n(w_{10},w_{20})$ i.i.d.\ with the distribution
\begin{equation*}
  \pr \Big\{U^n(w_{10},w_{20})=u^n\Big\}=\prod_{t=1}^nP(u_t).
\end{equation*}
Let $E(u^n)$ be the event $\{U^n(w_{10},w_{20})=u^n\}$.
Given $E(u^n)$, for every 
$(w_{id},z_i)\in [2^{nR_{id}}]\times [2^{nC_{id}}]$,
generate $V_i^{n}(w_{id},z_i|u^n)$ according to 
\begin{equation*}
\pr\Big\{V_i^{n}(w_{id},z_i|u^n)=v_i^n
\Big|E(u^n)\Big\}
=\prod_{t=1}^nP(v_{it}|u_t),
\end{equation*}
for $i=1,2$, where $P(v_1|u)$ and $P(v_2|u)$ are marginals of $P(v_1,v_2|u)$.

Fix $(w_{10},w_{20},w_{1d},w_{2d})$ and functions 
\begin{equation*}
  \nu_i:\big[2^{nC_{id}}\big] \rightarrow \mathcal{V}_i^n
\end{equation*}
for $i=1,2$. Let $E(u^n,\nu_1,\nu_2)$
denote the event where $U^n(w_{10},w_{20})=u^n$
and $V_1^{n}(w_{1d},.|u^n)=\nu_1(.)$, 
and $V_2^{n}(w_{2d},.|u^n)=\nu_2(.)$.
In addition,
for any $u^n,\nu_1$, and $\nu_2$, define the set 
\begin{equation*}
  \mathcal{A}(u^n,\nu_1,\nu_2):=
  \Big\{(z_1,z_2):(u^n,\nu_1(z_1),\nu_2(z_2))\in A_\delta^{(n)}\Big\}.
\end{equation*}
Given $E(u^n,\nu_1,\nu_2)$, if $\mathcal{A}(u^n,\nu_1,\nu_2)$
is nonempty, define 
\begin{equation*}
  \big(Z_1(u^n,\nu_1,\nu_2),Z_2(u^n,\nu_1,\nu_2)\big)
\end{equation*}
as a random pair that is uniformly distributed on  
$\mathcal{A}(u^n,\nu_1,\nu_2)$. Otherwise, set
$Z_i(u^n,\nu_1,\nu_2) = 1$ for $i=1,2$. 

Next, fix $(w_{10},w_{20},w_{1d},w_{2d})$ and let $E(u^n,v_1^n,v_2^n)$
denote the event where $U^n(w_{10},w_{20})=u^n$, $V_1^{n}(w_{1d},Z_1|u^n)=v_1^n$
and $V_2^{n}(w_{2d},Z_2|u^n)=v_2^n$. 
For every $w_{11}$ and $w_{22}$, generate the codewords 
$X_1^n(w_{11}|u^n,v_1^n)$ and $X_2^n(w_{22}|u^n,v_2^n)$ independently
according to the distributions
\begin{equation*}
  \pr\Big\{X_i^n(w_{ii}|u^n,v_i^n)=x_i^n
  \Big|E(u^n,v_1^n,v_2^n)\Big\}
  =\prod_{t=1}^nP(x_{it}|u_t,v_{it})
\end{equation*}
for $i=1,2$. This completes our codebook construction.

We next describe the encoding and decoding operations. Suppose
$W_1=(w_{10},w_{1d},w_{11})$ and $W_2=(w_{20},w_{2d},w_{22})$.
Encoders 1 and 2 send the pairs $(w_{10},w_{1d})$ and 
$(w_{20},w_{2d})$, respectively, to the cooperation facilitator.
Thus for $i=1,2$, $\varphi_i(w_i) = (w_{i0},w_{id})$.
The cooperation facilitator then transmits 
\begin{align*}
  \psi_1\big(\varphi_1(w_1),\varphi_2(w_2)\big) &= (w_{20}, Z_1)\\
  \psi_2\big(\varphi_1(w_1),\varphi_2(w_2)\big) &= (w_{10}, Z_2),
\end{align*}
to encoders 1 and 2, respectively. 

Using its knowledge of $(w_1,w_{20},Z_1)$, encoder 1
uses the $(U^n,V_1^n)$-codebook to transmit
$X_1^n(w_{11}|U^n,V_1^n)$. Similarly, using knowledge 
obtained from the cooperation facilitator, encoder 2
transmits $X_2^n(w_{22}|U^n,V_2^n)$. 

The decoder uses joint typicality decoding. Upon 
receiving $Y^n$ the decoder looks for a unique 
pair $(w_1,w_2)$ such that
\begin{align} \label{eq:typical}
  \MoveEqLeft \Big(U^n(w_{10},w_{20}), V_1^n(w_{1d},Z_1), V_2^n(w_{2d},Z_2),\notag\\
  &X_1^n(w_{11}),X_2^n(w_{22}), Y^n\Big)\in A_\epsilon^{(n)}.
\end{align}
If such a $(w_1,w_2)$ doesn't exist or exists but is not unique, 
the decoder declares an error.

\section{Error Analysis} \label{sec:error}
In this section, we study the achievability scheme more closely
and provide sufficient conditions for $(R_1,R_2)$ such that the probability of
error goes to zero. This immediately leads to Theorem \ref{thm:main}
which characterizes an achievable rate region for the MAC
with transmitter cooperation. 

Suppose the message pair $(w_1,w_2)$ is transmitted, where
$w_i=(w_{i0},w_{id},w_{ii})$.
If $(w_1,w_2)$ is the unique pair that satisfies Equation (\ref{eq:typical})
then there is no error. If such a pair does not exist or is not unique,
an error occurs. We denote this event by $\mathcal{E}$. Since directly finding an
upper bound on $\pr(\mathcal{E})$ is not straightforward, we upper bound $\mathcal{E}$
by the union of a finite number of events and then apply the union bound.
We give detailed proofs of the bounds mentioned in this section in 
Appendix \ref{app:error}.

In what follows, we denote $U^n(w_{10},w_{20})$ and
$V_i^n(w_{id},.|U^n)$ by $U^n$ and $V_i^n(.)$, respectively. 
In addition, we define
\begin{equation*}
  X_i^n(.)=X_i^n\big(w_{ii}|U^n,V_i^n(.)\big).
\end{equation*}
We denote instances of $V_i^n(.)$ and $X_i^n(.)$ with
$\nu_i(.)$ and $\chi_i(.)$, respectively. 
We also write $V_i^n$ and $X_i^n$
instead of $V_i^n(w_{id},Z_i|U^n)$ and $X_i^n(w_{ii}|U^n,V_i^n)$.

We denote the output of the decoder with $(\hat{w}_1,\hat{w}_2)$.
We denote $U^n(\hat{w}_{10},\hat{w}_{20})$ with
$\hat{U}^n$ and similarly define $\hat{V}_i^n$ and $\hat{X}_i^n$ 
for $i=1,2$.

We next describe the error events. First,
define $\mathcal{E}_0$ as
\begin{equation} \label{eq:e0}
  \mathcal{E}_0 = \Big\{
  (U^n,V_1^n,V_2^n)\notin A_\delta^{(n)}\Big\}.
\end{equation}
When $\mathcal{E}_0$ does not occur, the CF transmits
$(w_{20},Z_1)$ and $(w_{10},Z_2)$ to encoders 1 and
2, respectively, which correspond to a jointly typical triple
$(U^n,V_1^n,V_2^n)$.
Using the Mutual Covering Lemma for weakly typical sets
(Appendix \ref{app:mutual}), we show that $\pr(\mathcal{E}_0)$ 
goes to zero if $\zeta>4\epsilon$,
where $\zeta$ is defined by Equation (\ref{eq:zeta}). 

Next, define $\mathcal{E}_1$ as
\begin{equation*}
  \mathcal{E}_1 = \Big\{
  (U^n,V_1^n,V_2^n,X_1^n,X_2^n,Y^n)
  \notin A_\epsilon^{(n)}\Big\}.
\end{equation*}
This is the event where the codewords of the transmitted
message pair are not jointly typical with the received 
output $Y^n$. Then we have
$\pr(\mathcal{E}_1\setminus\mathcal{E}_0)\rightarrow 0$
as $n\rightarrow \infty$ if $\zeta<\Theta(\epsilon)-4\delta$.

If an error occurs and $\mathcal{E}_1^c$ holds, 
there must exist a message pair $(\hat{w}_1,\hat{w}_2)$ different 
from $(w_1,w_2)$ that satisfies (\ref{eq:typical}).
The message pair $(\hat{w}_1,\hat{w}_2)$, where
$\hat{w}_i=(\hat{w}_{i0},\hat{w}_{id},\hat{w}_{ii})$, 
may have $(\hat{w}_{10},\hat{w}_{20})\neq(w_{10},w_{20})$ or
$(\hat{w}_{10},\hat{w}_{20})=(w_{10},w_{20})$.

Define $\mathcal{E}_U$ as the event where
$(\hat{w}_{10},\hat{w}_{20})\neq(w_{10},w_{20})$.
In this case,
$(\hat{U}^n,\hat{V}_1^n,\hat{V}_2^n,\hat{X}_1^n,\hat{X}_2^n)$
and $Y^n$ are independent, which implies that $\pr(\mathcal{E}_U)$
goes to zero if $R_1+R_2<I(X_1,X_2;Y)-\zeta-7\epsilon$. 

If $(\hat{w}_{10},\hat{w}_{20})=(w_{10},w_{20})$, 
then either $(\hat{w}_{1d},\hat{w}_{2d})\neq(w_{1d},w_{2d})$ or 
$(\hat{w}_{1d},\hat{w}_{2d})=(w_{1d},w_{2d})$. If 
$(\hat{w}_{1d},\hat{w}_{2d})\neq(w_{1d},w_{2d})$, then
$\hat{w}_{1d}\neq w_{1d}$ but $\hat{w}_{2d}=w_{2d}$, or
$\hat{w}_{2d}\neq w_{2d}$ but $\hat{w}_{1d}=w_{1d}$, or $\hat{w}_{1d}\neq w_{1d}$
\emph{and} $\hat{w}_{2d}\neq w_{2d}$. 

Let $(i,j)\in\{(1,2),(2,1)\}$. If $\hat{w}_{id}\neq w_{id}$ and $\hat{w}_{jd}=w_{jd}$,
we may have $\hat{w}_{jj}\neq w_{jj}$ or $\hat{w}_{jj}=w_{jj}$.
We denote the former event by $\mathcal{E}_{V_iX_j}$ and the
latter by $\mathcal{E}_{V_i}$. 
Finally, we denote the event where 
$\hat{w}_{1d}\neq w_{1d}$ and $\hat{w}_{2d}\neq w_{2d}$
with $\mathcal{E}_{V_1V_2}$. 

For $(i,j)\in\{(1,2),(2,1)\}$, when $\mathcal{E}_{V_iX_j}$ occurs, 
$(\hat{V}_1^n,\hat{V}_2^n,\hat{X}_1^n,\hat{X}_2^n)$ and $Y^n$
are independent given $(U^n,V_j^n(.))$. 
This implies $\pr(\mathcal{E}_{V_iX_j})\rightarrow 0$ if
$(R_i-R_{i0})+R_{jj}< I(X_1,X_2;Y|U,V_j)-\zeta-8\epsilon$.

For $(i,j)\in\{(1,2),(2,1)\}$, when $\mathcal{E}_{V_i}$ occurs, we show that
$(\hat{V}_1^n,\hat{V}_2^n,\hat{X}_1^n,\hat{X}_2^n)$ and $Y^n$
are independent given $(U^n,V_j^n(.),X_j^n(.))$. 
This implies $\pr(\mathcal{E}_{V_i})\rightarrow 0$ if
$ R_i-R_{i0}< I(X_i;Y|U,V_j,X_j)-\zeta-8\epsilon$.

If $\mathcal{E}_{V_1V_2}$ occurs, 
$(\hat{V}_1^n,\hat{V}_2^n,\hat{X}_1^n,\hat{X}_2^n)$ and $Y^n$
are independent given $U^n$. Thus $\pr(\mathcal{E}_{V_1V_2})$
goes to zero if 
\begin{equation*} 
  (R_1-R_{10})+(R_2-R_{20})<I(X_1,X_2;Y|U)-\zeta-8\epsilon.
\end{equation*}

Finally, if an error occurs and 
the message pairs have the same $(w_{10},w_{20})$ and 
the same $(w_{1d},w_{2d})$, they must have different 
$(w_{11},w_{22})$. We define the events 
$\mathcal{E}_{X_i}$ and 
$\mathcal{E}_{X_1X_2}$ similarly to the events
for $(w_{1d},w_{2d})$. The relations
\begin{align*}
  \hat{X}_i^n &\rightarrow (U^n,V_1^n,V_2^n,X_j^n)
  \rightarrow Y^n\\
  (\hat{X}_1^n,\hat{X}_2^n)&\rightarrow (U^n,V_1^n,V_2^n)
  \rightarrow Y^n,
\end{align*}
hold for the events $\mathcal{E}_{X_i}$ and 
$\mathcal{E}_{X_1X_2}$, respectively. From these relations
it follows that
\begin{align*}
  \pr(\mathcal{E}_{X_i})\rightarrow 0 &\text{ if }
	R_{ii}< I(X_i;Y|U,V_1,V_2,X_j)-4\epsilon\\
	\pr(\mathcal{E}_{X_1X_2})\rightarrow 0 &\text{ if }
	R_{11}+R_{22}< I(X_1,X_2;Y|U,V_1,V_2)-4\epsilon.
\end{align*}
Not surprisingly, these bounds closely resemble the bounds
that appear in the capacity region of the classical MAC.

The bounds given in this section can be simplified further by replacing $R_i-R_{i0}$
and $R_{ii}$ with $(R_i-C_{i0})^+$ and $(R_i-C_i^\mathrm{in})^+$,
respectively, and noting that the set of 
all $(x,y)$ that satisfy $(x-a)^++(y-b)^+< c$
is the same as the set of all $(x,y)$ that satisfy 
$x-a<c$, $y-b<c$, and $(x-a)+(y-b)<c$.

Note that the general error event $\mathcal{E}$
is a subset of the union of the error events defined above. 
Thus if we apply the union bound and
choose $\delta$, $\epsilon$, and $\zeta$ to be arbitrarily small, 
we obtain Theorem \ref{thm:main}. 

\section{The Gaussian MAC} \label{sec:gaussian}
The Gaussian MAC~\cite{Wyner,Cover} is defined as the channel 
$Y_t = X_{1t}+X_{2t}+Z_t$, where $\{Z_t\}_{t=1}^n$ is an i.i.d.\ 
Gaussian process independent of $(X_1^n,X_2^n)$ and each
$Z_t$ is a Gaussian random variable with mean zero and variance $N$. 
In addition, the output power of encoder $i$ is constrained
by $P_i$, that is, $\sum_{t=1}^n x_{it}^2\leq nP_i$,
where $x_{it}$ is the output of encoder $i$ at time $t$ for $i=1,2$. 

For the Gaussian MAC, we modify the definition
of an achievable rate pair by adding the encoder power constraints to the
definition of the $(n,M_1,M_2)$ code for a MAC with a CF. Then 
the rate region 
$\mathscr{R}_\mathrm{mod}$
is achievable for the Gaussian MAC, where $\mathscr{R}_\mathrm{mod}$
is the same as $\mathscr{R}$ (Theorem \ref{thm:main}) with the additional constraints
$\mathbb{E}\big[X_i^2\big]\leq P_i$ for $i=1,2$
on the input distribution $P(u,v_1,v_2)P(x_1|u,v_1)P(x_2|u,v_2)$. 
This follows by replacing entropies with differential entropies and including
the input power constraints in the definition of $A_\epsilon^{(n)}$. This is 
possible since we use weakly typical sets \cite[p. 521]{CoverThomas} 
(rather than strongly typical sets) in the proof of Theorem \ref{thm:main}. 

If, in the calculation of $\mathscr{R}_\mathrm{mod}$, we limit ourselves
only to Gaussian input distributions, we get a rate region which we denote
by $\mathscr{R}_\mathrm{G}$. Note that $\mathscr{R}_\mathrm{G}$ is an inner bound
for the capacity region of a Gaussian MAC with a CF. We denote
the signal to noise ratio of encoder $i$ with $\gamma_i=\frac{P_i}{N}$ and define
$\bar{\gamma}=\sqrt{\gamma_1\gamma_2}$. The rate region 
$\mathscr{R}_\mathrm{G}$ is given by the next theorem. 
\begin{theorem} \label{thm:gaussian}
For the Gaussian MAC with a
$(\mathbf{C}_\mathrm{in},\mathbf{C}_\mathrm{out})$ CF, the achievable
rate region $\mathscr{R}_\mathrm{G}$ is given by the set of all
rate pairs $(R_1,R_2)$ that for $\{i,j\}=\{1,2\}$ satisfy
\begin{align*}
  R_i &< \frac{1}{2}\log(1+\rho_{ii}^2\gamma_i)+C_i^\mathrm{in}\\
	R_i &< \frac{1}{2}\log(1+\tilde{\rho}_{ii}^2\gamma_i)+C_{i0}\\
	R_1+R_2
	&< \frac{1}{2}\log(1+\rho_{11}^2\gamma_1+\rho_{22}^2\gamma_2)
	+C_1^\mathrm{in}+C_2^\mathrm{in}\\
	R_1+R_2 &< \frac{1}{2}\log(1+\rho_{ii}^2\gamma_i+\tilde{\rho}_{jj}^2\gamma_j)
	+C_i^\mathrm{in}+C_{j0}\\
	R_1+R_2
  &< \frac{1}{2}\log\big(1+(1-\rho_{10}^2)\gamma_1+(1-\rho_{20}^2)\gamma_2\\
	&\phantom{<\frac{1}{2}\log\big(}
	+2\rho_0\rho_{1d}\rho_{2d}\bar{\gamma}\big)+C_{10}+C_{20}\\
  R_1+R_2 &< \frac{1}{2}\log\big(1+\gamma_1+\gamma_2
	+2(\rho_{10}\rho_{20}+\rho_0\rho_{1d}\rho_{2d})\bar{\gamma}\big)
\end{align*}
for some $\rho_{10},\rho_{20},\rho_{1d},\rho_{2d}\in [0,1]$, 
and nonnegative constants $C_{10}$ and $C_{20}$
that satisfy Equation (\ref{eq:ci0}). 
In the above inequalities $\rho_0$, $\rho_{ii}$, and
$\tilde{\rho}_{ii}$ (for $i=1,2$) 
are given by
\begin{align}
  \frac{1}{2}\log\frac{1}{1-\rho_0^2} &\leq 
	(C_1^\mathrm{out}-C_{20})+(C_2^\mathrm{out}-C_{10})\nonumber\\
	\rho_{ii}^2 &= 1-\rho_{i0}^2-\rho_{id}^2 \label{eq:rii}\\
	\tilde{\rho}_{ii}^2 &= 1-\rho_{i0}^2-\rho_0^2\rho_{id}^2. \nonumber
\end{align}
\end{theorem}
\begin{figure} 
	\begin{center}
		\includegraphics[scale=0.35]{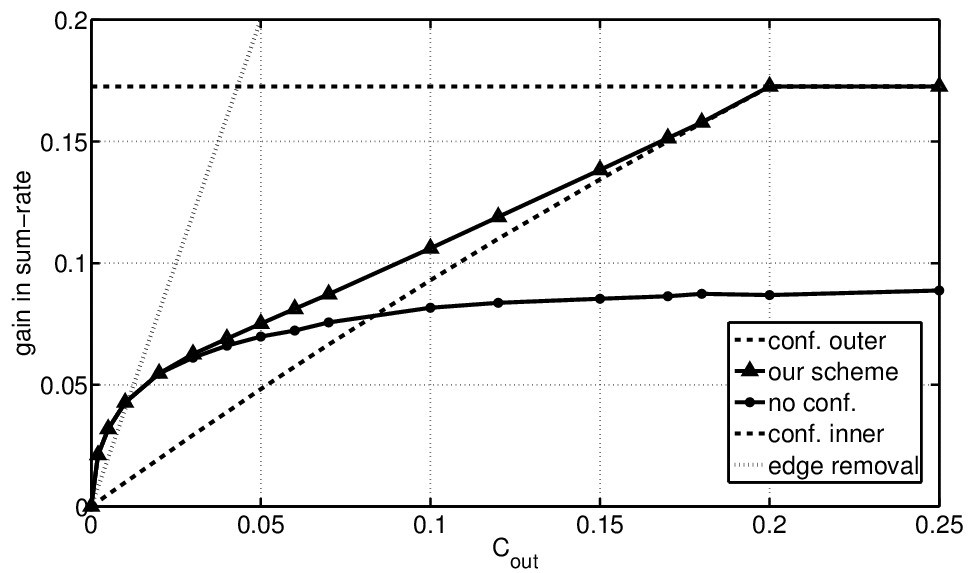} 
		\caption{The plot of the maximum sum-rate gain 
		by achieved by our scheme for the Gaussian MAC 
		with $\gamma_1=\gamma_2=10^3$ and $C^\mathrm{in}_1=C^\mathrm{in}_2=0.2$ 
		as a function of $C_\mathrm{out}$.} \label{fig:gaussian}
	\end{center}
\end{figure} 
We prove Theorem \ref{thm:gaussian} in Appendix \ref{app:gaussian} 
using techniques similar to \cite{Wigger}, in which the capacity region
of the Gaussian MAC with conferencing encoders is given.

Using Theorem \ref{thm:gaussian}, we can calculate the maximum
sum-rate of our scheme for the Gaussian MAC. We define the 
``sum-rate gain'' of a cooperation scheme as the difference
between the maximum sum-rate of that scheme and the maximum sum-rate
of the classical MAC scheme. In Figure 
\ref{fig:gaussian}, we plot the sum-rate gain of our scheme 
as a function of $C_1^\mathrm{out}=C_2^\mathrm{out}=:C_\mathrm{out}$
for $\gamma_1=\gamma_2=10^3$, $C_1^\mathrm{in}=C_2^\mathrm{in}=0.2$ and 
$C_\mathrm{out}\in [0,0.25]$. 
We also plot the conferencing
bounds in addition to the no conferencing sum-rate, which is the sum-rate
corresponding to a scheme that splits the rate between the coordination
and the classical MAC strategies and does
not make use of conferencing ($C_{10}=C_{20}=0$). 

Note that for any value of $C_\mathrm{out}$ for which the gain in sum-rate
is greater than $4C_\mathrm{out}$, adding a $(C_\mathrm{in},C_\mathrm{out})$-CF to
the Gaussian MAC results in a network that does not satisfy the edge removal 
property. The reason is that if we remove the output edges of the 
$(C_\mathrm{in},C_\mathrm{out})$-CF, the decrease in sum-capacity is greater than $4C_\mathrm{out}$,
which implies the decrease in either $R_1$ or $R_2$ (or both) is greater than
$2C_\mathrm{out}$, which is the total capacity of the removed edges.
On the plot, these are the points on our curve which fall 
above the ``edge removal line'', that is, the line whose equation is
given by $\mathrm{gain}=4C_\mathrm{out}$.

As we see, the scheme that makes no use of conferencing
performs well when $C_\mathrm{out}\ll C_\mathrm{in}$, 
and the conferencing scheme works well when $C_\mathrm{out}$ is close to $C_\mathrm{in}$
(and is optimal when $C_\mathrm{out}\geq C_\mathrm{in}$). Thus both strategies 
are necessary for our scheme to perform well over 
the entire range of  $C_\mathrm{out}$. 
In this case study, the maximum sum-rate of $\mathscr{R}_\mathrm{G}$ 
could have been obtained by a carefully designed 
time sharing between encoders which only cooperate through conferencing 
and encoders that use our scheme without conferencing. 
Whether this is representative of our scheme in 
general (specifically of $\mathscr{R}_\mathrm{mod}$ and $\mathscr{R}$) 
is subject to future research.

\section{Conclusion}
We study the cost and benefit of cooperation under a general model 
introduced in this paper. By adapting the coding strategy of 
Marton \cite{Marton} for the broadcast channel
to the setting of the MAC with cooperating encoders, we suggest a cooperation scheme
that combines the ideas of \cite{Marton} and \cite{ElGamalMeulen} with the conferencing 
strategy of Willems \cite{Willems}. Based on this scheme, we present an
inner bound for the MAC with a CF, which is sufficient
to show a large gain in sum-capacity as a result of transmitter cooperation.

Throughout the proof of our inner bound, we only make use of 
weakly typical sets \cite[p. 521]{CoverThomas}
rather than strongly typical sets \cite[p. 30]{ElGamalKim}. 
This allows the proof of our achievability result to go through for the Gaussian MAC 
without the use of quantization. In particular,
we present a proof (Appendix \ref{app:mutual}) of the 
Mutual Covering Lemma \cite[p. 208]{ElGamalKim} for weakly typical sets.

\section*{Acknowledgements}
This material is based upon work supported by the National Science Foundation 
under Grant No. CCF-1321129. The first author thanks Ming Fai Wong, 
Wei Mao, and Siddharth Jain for useful discussions. 

\appendices
\section{The Conferencing Encoders Model} \label{app:conf}

In the conferencing encoders model, introduced by Willems \cite{Willems},
each encoder sends partial information
regarding its message to the other encoder via a noiseless link. The capacity
of the links going from encoder 1 to encoder 2 and back are denoted by 
$C_{12}$ and $C_{21}$, respectively. At every time step, each encoder
sends information to the other encoder that is a function of its 
own message and what it received from that encoder during the previous
time steps. After this ``conference'' is over, each encoder transmits a codeword
over the channel that is a function of its message and information it received
during the conference. For a blocklength $n$ code, the amount of information
going from encoder 1 to encoder 2 and going back is bounded by $nC_{12}$ and
$nC_{21}$ bits, respectively. 

Even though the conference can go on for any finite number of steps, 
the achievability and converse results of Willems \cite{Willems} demonstrate 
that a single step of conferencing suffices to achieve capacity. In one-step conferencing,
encoder 1 sends a rate $C_{12}$ function of its message to encoder 2 and 
encoder 2 sends a rate $C_{21}$ function of its message to encoder 1. Then the encoders
treat the shared messages as a single rate $C_{12}+C_{21}$ common message and use the
channel coding strategy of Slepian and Wolf \cite{SlepianWolf}.

We denote the capacity of a MAC 
$(\mathcal{X}_1\times\mathcal{X}_2,P(y|x_1,x_2),\mathcal{Y})$
with a $(C_{12},C_{21})$ conference with
$\mathscr{C}_\mathrm{conf}(C_{12},C_{21})$, which is given by 
the set of all rate pairs $(R_1,R_2)$ that satisfy
\begin{equation} \label{eq:confcap}
\begin{aligned}
 R_1 &\leq I(X_1;Y|X_2,U)+C_{12}\\
 R_2 &\leq I(X_2;Y|X_1,U)+C_{21}\\
 R_1+R_2 &\leq I(X_1,X_2;Y|U)+C_{12}+C_{21}\\
 R_1+R_2 &\leq I(X_1,X_2;Y)
\end{aligned}
\end{equation}
for some distribution $P(u)P(x_1|u)P(x_2|u)$. 

Note that the achievable rate region
$\mathscr{R}(\mathbf{C}^\mathrm{in},\mathbf{C}^\mathrm{out})$
introduced in Section \ref{sec:coop}
satisfies the conferencing bounds. 
For the inner bound, if we choose $|\mathcal{V}_1|=|\mathcal{V}_2|=1$,
and
\begin{align*}
  C_{10} &= \min\big\{C_1^\mathrm{in},C_2^\mathrm{out}\big\}\\
  C_{20} &= \min\big\{C_2^\mathrm{in},C_1^\mathrm{out}\big\},
\end{align*} 
we see that the bounds in the definition of $\mathscr{R}$ simplify
to those given by (\ref{eq:confcap}) for $C_{12}=C_{10}$ and $C_{21}=C_{20}$. 
For the outer bound, notice that the inequalities 
\begin{align*}
  R_1 &< I(X_1;Y|U,V_1,V_2,X_2)+C_1^\mathrm{in}\\
	R_2 &< I(X_2;Y|U,V_1,V_2,X_1)+C_2^\mathrm{in}\\
	R_1+R_2 &< I(X_1,X_2;Y|U,V_1,V_2)
	+C_1^\mathrm{in}+C_2^\mathrm{in} \\
	R_1+R_2 &< I(X_1,X_2;Y),
\end{align*}
which appear in the definition of $\mathscr{R}$ 
are the same as those in (\ref{eq:confcap}) 
for $C_{12}=C_1^\mathrm{in}$ and $C_{21}=C_2^\mathrm{in}$,
since $X_1$ and $X_2$ are independent given $(U,V_1,V_2)$.

\section{Proof of Theorem \ref{thm:growthrate}} \label{app:growthrate}

Fix $\mathbf{C}_\mathrm{in}=(C_1^\mathrm{in},C_2^\mathrm{in})$ 
for some positive $C_1^\mathrm{in}$ and $C_2^\mathrm{in}$ 
and let $\mathbf{C}_\mathrm{out}=(C_\mathrm{out},C_\mathrm{out})$. 
Define $g:\mathbb{R}_{\geq 0}\rightarrow \mathbb{R}_{\geq 0}$
as 
\begin{equation*}
  g(C_\mathrm{out})=
	C_\mathrm{sum}(\mathbf{C}_\mathrm{in},\mathbf{C}_\mathrm{out}).
\end{equation*}
Then by Theorem \ref{thm:main}, $g(C_\mathrm{out})$,
is bounded from below by the maximum of
\begin{equation}
\begin{aligned} \label{eq:sumcapacity}
  \min
	\big\{&I(X_1,X_2;Y|U,V_1,V_2)+C_1^\mathrm{in}+C_2^\mathrm{in},\\
	&I(X_1,X_2;Y|U,V_1)+C_1^\mathrm{in}+C_{20},\\
	&I(X_1,X_2;Y|U,V_2)+C_{10}+C_2^\mathrm{in},\\
	&I(X_1;Y|U,V_2,X_2)+I(X_2;Y|U,V_1,X_1)+C_{10}+C_{20},\\
	&I(X_1,X_2;Y|U)+C_{10}+C_{20},I(X_1,X_2;Y)\big\}
\end{aligned}
\end{equation}
calculated over all alphabets $(\mathcal{U},\mathcal{V}_1,\mathcal{V}_2)$, 
all nonnegative constants $(C_{10},C_{20})$ 
satisfying Equation (\ref{eq:ci0}), and all probability distributions
$P(u,v_1,v_2)P(x_1|u,v_1)P(x_2|u,v_2)$
that satisfy 
\begin{equation*}
  I(V_1;V_2|U)\leq 2C_\mathrm{out}-C_{10}-C_{20}.
\end{equation*}

We next find a simpler lower bound for $g(C_\mathrm{out})$ by evaluating the minimum
in Equation (\ref{eq:sumcapacity}) for fixed alphabets, constants, and a special
family of distributions. To this end, choose the sets $\mathcal{U}$, $\mathcal{V}_1$,
and $\mathcal{V}_2$ such that  $|\mathcal{U}|=1$ and 
$\mathcal{X}_i\subseteq \mathcal{V}_i$ for $i=1,2$. 
In addition, let $C_{10}=C_{20}=0$ and 
let $P_a(x_1)P_a(x_2)$ and $P_b(x_1,x_2)$ be distributions such 
that 
\begin{align*}
  I_a(X_1,X_2;Y) &= \max_{P(x_1)P(x_2)}I(X_1,X_2;Y)\\
	I_b(X_1,X_2;Y) &> I_a(X_1,X_2;Y).
\end{align*}
Fix $(v_1^*,v_2^*)\in \mathcal{V}_1\times \mathcal{V}_2$.
For every $\lambda\in [0,1]$, define
\begin{align*}
  \MoveEqLeft P_\lambda(v_1,v_2,x_1,x_2)\\&=
	(1-\lambda)\mathbf{1}\{v_1=v_1^*\}\mathbf{1}\{v_2=v_2^*\}P_a(x_1)P_a(x_2)\\
	&\phantom{=}+\lambda P_b(v_1,v_2)\mathbf{1}\{x_1=v_1\}\mathbf{1}\{x_2=v_2\}.
\end{align*}
Fix $\epsilon>0$. Consider the equation
\begin{equation*}
  I_{\lambda^*}(V_1;V_2)+2\epsilon\lambda^*=2C_\mathrm{out}. 
\end{equation*}
By Lemma \ref{lem:derivative} (see end of appendix), 
\begin{equation*}
  \frac{dC_\mathrm{out}}{d\lambda^*}\Big|_{\lambda^*=0^+}
	= \epsilon>0.
\end{equation*}
Thus by the inverse function theorem, there 
exists a continuous increasing
function $\lambda^*=\lambda^*(C_\mathrm{out})$ on $[0,\delta_1)$
for some $\delta_1>0$. Thus for $C_\mathrm{out}<\delta_1$, 
$g(C_\mathrm{out})$ is bounded from below by
\begin{align*}
  \min
	\big\{&I_{\lambda^*}(X_1,X_2;Y|V_1,V_2)+C_1^\mathrm{in}+C_2^\mathrm{in},\\
	&I_{\lambda^*}(X_1,X_2;Y|V_1)+C_1^\mathrm{in},I_{\lambda^*}(X_1,X_2;Y|V_2)+C_2^\mathrm{in},\\
	&I_{\lambda^*}(X_1;Y|V_2,X_2)+I_{\lambda^*}(X_2;Y|V_1,X_1),\\
	&I_{\lambda^*}(X_1,X_2;Y)\big\}.
\end{align*}
One of the terms that appears in the above lower bound is 
$I_{\lambda^*}(X_1;Y|V_2,X_2)+I_{\lambda^*}(X_2;Y|V_1,X_1)$. 
We can further bound this expression using the next lemma. 
\begin{lemma}\label{lem:trivial}
For any memoryless MAC,
\begin{equation*}
  I(X_1;Y|X_2)+I(X_2;Y|X_1)\geq I(X_1,X_2;Y)-I(X_1;X_2).
\end{equation*}
\end{lemma}
\begin{IEEEproof} We have
\begin{align*}
  \MoveEqLeft I(X_1;Y|X_2)+I(X_2;Y|X_1)+I(X_1;X_2)\\
	&= I(X_1;Y|X_2)+I(X_2;X_1,Y)\\
	&= I(X_1;Y|X_2)+I(X_2;Y)+I(X_1;X_2|Y)\\
	&= I(X_1,X_2;Y)+I(X_1;X_2|Y).
\end{align*}
The result of the lemma now follows from the nonnegativity
of mutual information.
\end{IEEEproof}
If in the above lemma we replace $X_i$ with $(V_i,X_i)$ with
arbitrary distribution $P(v_1,v_2)P(x_1|v_1)P(x_2|v_2)$ and
simplify we get 
\begin{equation*}
  I(X_1;Y|V_2,X_2)+I(X_2;Y|V_1,X_1)\geq I(X_1,X_2;Y)-I(V_1;V_2),
\end{equation*}
since $(V_1,V_2)\rightarrow (X_1,X_2)\rightarrow Y$ is a Markov
chain and 
\begin{align*}
  \MoveEqLeft I(V_1,X_1;V_2,X_2)\\ 
	&= H(V_1,X_1)+H(V_2,X_2)-H(V_1,V_2,X_1,X_2)\\
	&= I(V_1;V_2).
\end{align*}
Therefore, for $C_\mathrm{out}<\delta_1$, $g(C_\mathrm{out})$ 
is bounded from below by 
\begin{equation*}
  \min_{i\in\{0,1,2,3\}} g_i(C_\mathrm{out}),
\end{equation*}	
where
\begin{align*}
  g_0(C_\mathrm{out})  &= I_{\lambda^*}(X_1,X_2;Y)-I_{\lambda^*}(V_1;V_2)\\
	g_1(C_\mathrm{out})  &= I_{\lambda^*}(X_1,X_2;Y|V_1)+C_1^\mathrm{in}\\
	g_2(C_\mathrm{out})  &= I_{\lambda^*}(X_1,X_2;Y|V_2)+C_2^\mathrm{in}\\
	g_3(C_\mathrm{out}) &= I_{\lambda^*}(X_1,X_2;Y|V_1,V_2)+C_1^\mathrm{in}+C_2^\mathrm{in}.
\end{align*}
Note that if $C_\mathrm{out}=0$, then $\lambda^*(C_\mathrm{out})=0$ and
\begin{equation*}
  \min_{i\in\{0,1,2,3\}} g_i(C_\mathrm{out})=g_{0}(0)=I_a(X_1,X_2;Y),
\end{equation*}
since 
\begin{equation*}
  P_0(v_1,v_2,x_1,x_2)
	= \mathbf{1}\{v_1=v_1^*\}\mathbf{1}\{v_2=v_2^*\}P_a(x_1)P_a(x_2)
\end{equation*}
and $\min\{C_1^\mathrm{in},C_2^\mathrm{in}\}>0$. 
Furthermore, as the $g_i$'s are continuous in $C_\mathrm{out}$, 
there exists a positive $\delta$ smaller than $\delta_1$ 
such that for every $C_\mathrm{out}<\delta$,
\begin{equation*}
  \min_{i\in\{0,1,2,3\}} g_i(C_\mathrm{out})=g_{0}(C_\mathrm{out}).
\end{equation*} 
Therefore, for $C_\mathrm{out}<\delta$, 
$g(C_\mathrm{out})$ is bounded from below by
\begin{equation*}
  g_0(C_\mathrm{out})=I_{\lambda^*}(X_1,X_2;Y)-I_{\lambda^*}(V_1;V_2).
\end{equation*}
Since, in addition,  
\begin{equation*}
  g(0)=g_0(0)=I_a(X_1,X_2;Y),
\end{equation*}
we have
\begin{align*}
  g'(0) &\geq g'_0(0)=\frac{dg_0}{dC_\mathrm{out}}\Big|_{C_\mathrm{out}=0^+}\\
	&= \frac{dg_0}{d\lambda^*}\Big|_{\lambda^*=0^+}
	\cdot \frac{d\lambda^*}{dC_\mathrm{out}}\Big|_{C_\mathrm{out}=0^+}\\
	&\geq \frac{1}{\epsilon}\big(I_b(X_1,X_2;Y)-I_a(X_1,X_2;Y)\big),
\end{align*}
where the last inequality follows from Lemma \ref{lem:derivative}
(see end of appendix).
Since $\epsilon$ can be chosen to be arbitrarily small, we 
must have $g'(0)=+\infty$.

In the special case where our channel is a Gaussian MAC, if
we choose $\rho_{10}=\rho_{20}=0$, $\rho_{1d}=\rho_{2d}=:\rho_d$,
and $C_{10}=C_{20}=0$ in Theorem \ref{thm:gaussian}, 
we see that $g(C_\mathrm{out})$ is bounded from below by the
maximum of 
\begin{equation*}
 \min_{0\leq i\leq 4}f_i(\rho_0,\rho_d),
\end{equation*}
calculated over $(\rho_0,\rho_d)$, where $\rho_d\in [0,1]$,
\begin{equation*}
  0 \leq \rho_0 \leq \sqrt{1-e^{-4C_\mathrm{out}}}=:\rho_0(C_\mathrm{out}).
\end{equation*}
The $f_i$'s are defined as 
\begin{align*}
  f_0(\rho_0,\rho_d) &=\frac{1}{2}\log(1+\gamma_1+\gamma_2+2\rho_0\rho_d^2\sqrt{\gamma_1\gamma_2})\\
	f_1(\rho_0,\rho_d) &=\frac{1}{2}\log\big(1+(1-\rho_0^2\rho_d^2)\gamma_1\big)\\
	&\phantom{=}+\frac{1}{2}\log\big(1+(1-\rho_0^2\rho_d^2)\gamma_2\big)\\
	f_2(\rho_0,\rho_d) &=\frac{1}{2}\log\big(1+(1-\rho_d^2)\gamma_1
	+(1-\rho_0^2\rho_d^2)\gamma_2\big)+C_1^\mathrm{in}\\
	f_3(\rho_0,\rho_d) &=\frac{1}{2}\log\big(1+(1-\rho_0^2\rho_d^2)\gamma_1
	+(1-\rho_d^2)\gamma_2\big)+C_2^\mathrm{in}\\
	f_4(\rho_0,\rho_d) &=\frac{1}{2}\log\big(1+(1-\rho_d^2)\gamma_1
	+(1-\rho_d^2)\gamma_2\big)+C_1^\mathrm{in}+C_2^\mathrm{in}.
\end{align*}
Next define the function $F(\rho_0,\rho_d^*)$ as
\begin{equation*}
  F(\rho_0,\rho_d)=f_0(\rho_0,\rho_d)-\min_{1\leq i\leq 4}f_i(\rho_0,\rho_d).
\end{equation*}
Note that $F(1,0)<0$ since $\gamma_1$, $\gamma_2$, $C_1^\mathrm{in}$,
and $C_2^\mathrm{in}$ are positive.
Since $F$ is continuous, there exists $\rho_d^*>0$
such that $F(1,\rho_d^*)<0$. However, for any $\rho_d$, $F(.,\rho_d)$ is an
increasing function of $\rho_0$. Thus for any $\rho_0\leq \rho_0(C_\mathrm{out})$,
\begin{equation*}
  F(\rho_0,\rho_d^*)\leq F(1,\rho_d^*) <0.
\end{equation*}
In particular, $F\big(\rho_0(C_\mathrm{out}),\rho\big)<0$. This implies
\begin{equation*}
  g(C_\mathrm{out})>f_0\big(\rho_0(C_\mathrm{out}),\rho_d^*\big).
\end{equation*}
To calculate $f_0\big(\rho_0(C_\mathrm{out}),\rho_d^*\big)$ we make use of the next lemma.
\begin{lemma}
For constants $a$ and $b$ ($b>0$) we have
\begin{equation*}
  \log\big(1+a\sqrt{1-e^{-bx}}\big)=a\sqrt{bx}+o(\sqrt{x})\text{ for }x>0.
\end{equation*}
\end{lemma}
\begin{IEEEproof}
We have
\begin{align*}
  \log\big(1+a\sqrt{1-e^{-bx}}\big) &=
	\log\big(1+a\sqrt{bx+o(x)}\big)\\
	&=\log\big(1+a\sqrt{bx}+o(\sqrt{x})\big)\\
	&= a\sqrt{bx}+o(\sqrt{x}).
\end{align*}
\end{IEEEproof}
By the previous lemma, 
\begin{align*}
  f_0\big(\rho_0(C_\mathrm{out}),\rho_d^*\big)-g(0)
	&= \frac{1}{2}\log\big(1+\frac{a}{2}\sqrt{1-e^{-4C_\mathrm{out}}}\big)\\
	&= a\sqrt{C_\mathrm{out}}+o\big(\sqrt{C_\mathrm{out}}\big),
\end{align*}
where $g(0) = \frac{1}{2}\log(1+\gamma_1+\gamma_2)$ and
\begin{equation*}
	a = \frac{4\rho_d^{*2}\sqrt{\gamma_1\gamma_2}}{1+\gamma_1+\gamma_2}.
\end{equation*}
To get the result stated in the theorem, it suffices to choose $\alpha$ such
that $0<\alpha<a$. The next lemma, used in the appendix to calculate the
derivatives of $I_\lambda(V_1;V_2)$ and $I_\lambda(X_1,X_2;Y)$, follows.  

\begin{lemma} \label{lem:derivative}
Let $P_0(x_1)P_0(x_2)$ and $P_1(x_1,x_2)$ be joint
distributions on $\mathcal{X}_1\times \mathcal{X}_2$. 
For every $\lambda\in [0,1]$, define the distribution
$P_\lambda(x_1,x_2)$ as
\begin{equation*}
  P_\lambda(x_1,x_2)=\lambda P_1(x_1,x_2)+(1-\lambda)P_0(x_1)P_0(x_2).
\end{equation*}
Then 
\begin{equation*}
  \frac{d}{d\lambda}I_\lambda(X_1;X_2)\Big|_{\lambda=0^+}=0.
\end{equation*}
Furthermore, if $P_\lambda(x_1,x_2,y)=P_\lambda(x_1,x_2)P(y|x_1,x_2)$,
then
\begin{equation*}
  \frac{d}{d\lambda}I_\lambda(X_1,X_2;Y)\Big|_{\lambda=0^+}
	\geq I_1(X_1,X_2;Y)-I_0(X_1,X_2;Y).
\end{equation*} 
\end{lemma} 
\begin{IEEEproof} Note that for every $(x_1,x_2)$,
\begin{equation*}
  \frac{d}{d\lambda}P_\lambda(x_1,x_2)=
	P_1(x_1,x_2)-P_0(x_1)P(x_2).
\end{equation*}
Since
\begin{equation*}
  I_\lambda(X_1;X_2)=\sum_{x_1,x_2}
	P_\lambda(x_1,x_2)\log
	\frac{P_\lambda(x_1,x_2)}{P_\lambda(x_1)P_\lambda(x_2)},
\end{equation*}
we have
\begin{align*}
  \MoveEqLeft \frac{d}{d\lambda} I_\lambda(X_1;X_2)\\
	&= \sum_{x_1,x_2}\big(P_1(x_1,x_2)-P_0(x_1)P_0(x_2)\big)
	\log \frac{P_\lambda(x_1,x_2)}{P_\lambda(x_1)P_\lambda(x_2)}\\
	&\phantom{=} +\sum_{x_1,x_2}P_\lambda(x_1,x_2)\Bigg(
	\frac{P_1(x_1,x_2)-P_0(x_1)P_0(x_2)}{P_\lambda(x_1,x_2)}\\
	&\phantom{=+\sum_{x_1,x_2}\Bigg(}
	-\frac{P_1(x_1)-P_0(x_1)}{P_\lambda(x_1)}
	-\frac{P_1(x_2)-P_0(x_2)}{P_\lambda(x_2)}\Bigg)\\
	&= \sum_{x_1,x_2}\big(P_1(x_1,x_2)-P_0(x_1)P_0(x_2)\big)
	\log \frac{P_\lambda(x_1,x_2)}{P_\lambda(x_1)P_\lambda(x_2)}.
\end{align*}
Thus 
\begin{equation*}
  \frac{d}{d\lambda}I_\lambda(X_1;X_2)\Big|_{\lambda=0^+}=0.
\end{equation*}

For the second part, we write
\begin{equation*}  
  I_\lambda(X_1,X_2;Y)=H_\lambda(Y)-H_\lambda(Y|X_1,X_2)
\end{equation*}
and calculate the derivatives of $H_\lambda(Y)$ and
$H_\lambda(Y|X_1,X_2)$ separately. Note that 
\begin{equation*}
  H_\lambda(Y)=-\sum_y P_\lambda(y)\log P_\lambda(y),
\end{equation*}
thus
\begin{align*}
  \frac{dH_\lambda(Y)}{d\lambda}
	&= -\sum_y \big(1+\log P_\lambda(y)\big)\big(P_1(y)-P_0(y)\big)\\
	&= \sum_y \big(P_0(y)-P_1(y)\big)\log P_\lambda(y)\\
	&= H_0(Y)+H_1(Y)\\
	&\phantom{=}+D\big(P_1(y)\|P_\lambda(y)\big)
	-D\big(P_0(y)\|P_\lambda(y)\big).
\end{align*}
Furthermore, we have 
\begin{equation*}
  H(Y|X_1,X_2)=\sum_{x_1,x_2}P_\lambda(x_1,x_2)H(Y|X_1=x_1,X_2=x_2),
\end{equation*}
so 
\begin{equation*}
  \frac{d}{d\lambda}H_\lambda(Y|X_1,X_2)
	= H_1(Y|X_1,X_2)-H_0(Y|X_1,X_2).
\end{equation*}
Therefore,
\begin{align*}
  \frac{d}{d\lambda}I_\lambda(X_1,X_2;Y)
	&= I_1(X_1,X_2;Y)-I_0(X_1,X_2;Y)\\
	&\phantom{=}+D\big(P_1(y)\|P_\lambda(y)\big)
	-D\big(P_0(y)\|P_\lambda(y)\big).
\end{align*}
Thus 
\begin{equation*}
  \frac{d}{d\lambda}I_\lambda(X_1,X_2;Y)\Big|_{\lambda=0^+}
	\geq I_1(X_1,X_2;Y)-I_0(X_1,X_2;Y).
\end{equation*} 
\end{IEEEproof}

\section{The $C_1^\mathrm{in}=C_2^\mathrm{in}=\infty$ Case}
\label{app:infty}

In this appendix, we find a simple representation for 
$\mathscr{R}(\mathbf{C}_\mathrm{in},\mathbf{C}_\mathrm{out})$
(Section \ref{sec:coop}) in the case where 
$C_1^\mathrm{in}=C_2^\mathrm{in}=\infty$. If we denote this 
region with $\mathscr{R}$, then $\mathscr{R}$ consists of all 
rate pairs $(R_1,R_2)$ that satisfy
\begin{align*}
  R_1 &< I(X_1;Y|U,V_2,X_2)+C_{10}\\
	R_2 &< I(X_2;Y|U,V_1,X_1)+C_{20}\\
	R_1+R_2 &< I(X_1,X_2;Y|U)+C_{10}+C_{20}\\
	R_1+R_2 &< I(X_1,X_2;Y),
\end{align*}
for some $C_{10}\leq C_2^\mathrm{out}$ and $C_{20}\leq C_1^\mathrm{out}$,
and some distribution $P(u,v_1,v_2)P(x_1|u,v_1)P(x_2|u,v_2)$ that satisfies
\begin{equation} \label{eq:cond11}
  I(V_1;V_2|U)\leq (C_1^\mathrm{out}-C_{20})+(C_2^\mathrm{out}-C_{10})
\end{equation}
Note that this region is contained in the region consisting of all rate pairs
$(R_1,R_2)$ that satisfy
\begin{align*}
  R_1 &< I(X_1;Y|U,X_2)+C_{10}\\
	R_2 &< I(X_2;Y|U,X_1)+C_{20}\\
	R_1+R_2 &< I(X_1,X_2;Y|U)+C_{10}+C_{20}\\
	R_1+R_2 &< I(X_1,X_2;Y),
\end{align*}
for some $C_{10}\leq C_2^\mathrm{out}$ and $C_{20}\leq C_1^\mathrm{out}$,
and some distribution $P(u,x_1,x_2)$ that satisfies
\begin{equation} \label{eq:cond22}
  I(X_1;X_2|U)\leq (C_1^\mathrm{out}-C_{20})+(C_2^\mathrm{out}-C_{10}).
\end{equation}
This follows from the fact that 
\begin{equation*}
 (U,V_1,V_2)\rightarrow (X_1,X_2)\rightarrow Y
\end{equation*}
is a Markov chain, and any distribution $P(u,v_1,v_2)P(x_1|u,v_1)P(x_2|u,v_2)$ 
that satisfies Equation (\ref{eq:cond11}) also satisfies Equation (\ref{eq:cond22}),
since
\begin{equation*}
  I(X_1;X_2|U) \leq I(V_1,X_1;V_2,X_2|U) = I(V_1;V_2|U).
\end{equation*}
To show that these two regions are in fact equal, it now suffices to choose
$\mathcal{V}_i=\mathcal{X}_i$ for $i=1,2$, and
\begin{align*}
  \MoveEqLeft
  P(u,v_1,v_2)P(x_1|u,v_1)P(x_2|u,v_2)\\
	&= P(u,v_1,v_2)\delta(x_1-v_1)\delta(x_2-v_2)
\end{align*}
in the definition of the first region. 

\section{Proof of Theorem \ref{thm:convexity}} \label{app:convexity}

Fix $\lambda\in (0,1)$. Suppose $(R_{1a},R_{2a})\in \mathscr{R}_a$
and $(R_{1b},R_{2b})\in \mathscr{R}_b$. Then there exist constants
$(C_{10}^a,C_{20}^a)$ and $(C_{10}^b,C_{20}^b)$
and input distributions
\begin{align*}
  &P_a(u)P_a(v_1,v_2|u)P_a(x_1|u,v_1)P_a(x_2|u,v_2)\\
  &P_b(u)P_b(v_1,v_2|u)P_b(x_1|u,v_1)P_b(x_2|u,v_2)
\end{align*}
that satisfy the constraints of $\mathscr{R}_a$ and $\mathscr{R}_b$
for the rate pairs $(R_{1a},R_{2a})$ and $(R_{1b},R_{2b})$, respectively.
We show that the rate
pair $(R_{1\lambda},R_{2\lambda})\in\mathscr{R}_\lambda$, where
for $i=1,2$,
\begin{equation*}
  R_{i\lambda}=\lambda R_{ia}+(1-\lambda)R_{ib}.
\end{equation*}
First define 
\begin{equation*}
  C^\lambda_{i0} = \lambda C^a_{i0} + (1-\lambda) C^b_{i0}
\end{equation*}
for $i=1,2$. Notice that by this definition, these constants satisfy
the constraints of $\mathscr{R}_\lambda$.
 
Next, consider the input distribution
\begin{equation} \label{eq:dist} 
  P(u')P(v_1,v_2|u')P(x_1|u',v_1)P(x_2|u',v_2)
\end{equation}
where $u'=(u,s)$, $s\in \{a,b\}$, $P(s=a)=\lambda$, 
\begin{equation*}
  P(u|s)= \begin{cases}
    P_a(u) &\text{if }s=a\\
    P_b(u) &\text{if }s=b,
    \end{cases}
\end{equation*}
and
\begin{equation*}
  P(v_1,v_2|u,s)= \begin{cases}
    P_a(v_1,v_2|u) &\text{if }s=a\\
    P_b(v_1,v_2|u) &\text{if }s=b.
    \end{cases}
\end{equation*}
Define $P(x_1|u',v_1)$ and $P(x_2|u',v_2)$ similarly. We show
that $(R_{1\lambda},R_{2\lambda})$ satisfies the bounds of 
$\mathscr{R}_\lambda$ for this input distribution.
Note that for any mutual information of the form $I(A;B|U,C)$ we have
\begin{equation*}
  I(A;B|U',C)=\lambda I_a(A;B|U,C) +(1-\lambda) I_b(A;B|U,C),
\end{equation*}
where $A$, $B$, and $C$ are arbitrary random variables.
Except for $I(X_1,X_2;Y)$, all the other mutual information terms appearing
in the definition of $\mathscr{R}_\lambda$ are of this form. 
If we compute $I(X_1,X_2;Y)$ with respect to (\ref{eq:dist}) we get
\begin{align*}
  I(X_1,X_2;Y) &\geq I(X_1,X_2;Y|S)\\
  &= \lambda I_a(X_1,X_2;Y) +(1-\lambda) I_b(X_1,X_2;Y),
\end{align*}
where the inequality holds since 
\begin{equation*}
  S\rightarrow (X_1,X_2)\rightarrow Y
\end{equation*}
is a Markov chain. Thus 
$(R_{1\lambda},R_{2\lambda})\in \mathscr{R}_\lambda$ and the 
proof is complete. 

\section{Details of Error Analysis} \label{app:error}
Before going into the proofs of the error bounds of Section \ref{sec:error}
we need to study the the distribution of our code in more detail.
Note that 
\begin{align*}
  \MoveEqLeft
  P_\mathrm{code}(u^n,\nu_1,\nu_2,v_1^n,v_2^n,x_1^n,x_2^n,y^n)\\
	&= P(u^n)P(\nu_1|u^n)P(\nu_2|u^n)P(v_1^n,v_2^n|u^n,\nu_1,\nu_2)\\
	&\phantom{=} \times P(x_1^n|u^n,v_1^n)P(x_2^n|u^n,v_2^n)P(y^n|x_1^n,x_2^n).
\end{align*}
The next lemma relates $P_\mathrm{code}(v_1^n,v_2^n|u^n)$ to the marginals
of $P(v_1^n,v_2^n|u^n)$, which is the distribution we use in the definition
of $A_\delta^{(n)}$ and $A_\epsilon^{(n)}$. 
\begin{lemma} \label{lem:pbounds} 
For all $(u^n,v_1^n,v_2^n)$,
\begin{equation*}
  P_\mathrm{code}(v_1^n,v_2^n|u^n)
	\leq 2^{n(C_{1d}+C_{2d})}P(v_1^n|u^n)P(v_2^n|u^n).
\end{equation*}
\end{lemma}
\begin{IEEEproof}
Note that
\begin{align} 
  \MoveEqLeft
  P_\mathrm{code}(v_1^n,v_2^n|u^n) \label{eq:pcode}\\
	&=\sum_{\nu_1,\nu_2}P(\nu_1|u^n)P(\nu_2|u^n) \nonumber 
	P(v_1^n,v_2^n|u^n,\nu_1,\nu_2). 
\end{align}
We have
\begin{equation*}
  P(v_1^n,v_2^n|u^n,\nu_1,\nu_2)
  \leq \mathbf{1}\big\{\nu_1^{-1}(v_1^n)\neq \emptyset\big\}
	\mathbf{1}\big\{\nu_2^{-1}(v_2^n)\neq \emptyset\big\},
\end{equation*}
where $\nu_i^{-1}(v_i^n)$, for $i=1,2$, is defined as 
\begin{equation*}
  \nu_i^{-1}(v_i^n)=\big\{z:\nu_i(z)=v_i^n\big\}.
\end{equation*}
We thus calculate, for $i=1,2$,
\begin{align*}
  \MoveEqLeft
  \sum_{\nu_i}P(\nu_i|u^n)\mathbf{1}\big\{\nu_i^{-1}(v_i^n)\neq \emptyset\big\}\\
  &= 1-\sum_{\nu_i}P(\nu_i|u^n)\mathbf{1}\big\{\nu_i^{-1}(v_i^n)=\emptyset\big\}\\
	&= 1-\big(1-P(v_i^n|u^n)\big)^{2^{nC_{id}}}\\
	&\leq 2^{nC_{id}}P(v_i^n|u^n),
\end{align*}
where the last inequality follows from the fact that 
$1-\alpha x\leq (1-x)^\alpha$ for all  
nonnegative $\alpha$ and $x$. Therefore,
\begin{align*}
  P_\mathrm{code}(v_1^n,v_2^n|u^n)
  &\leq \sum_{\nu_1}P(\nu_1|u^n)\mathbf{1}\big\{\nu_1^{-1}(v_1^n)\neq \emptyset\big\}\\
	&\phantom{=}\times
	\sum_{\nu_2}P(\nu_2|u^n)\mathbf{1}\big\{\nu_2^{-1}(v_2^n)\neq \emptyset\big\}\\
  &\leq 2^{n(C_{1d}+C_{2d})}P(v_1^n|u^n)P(v_2^n|u^n),
\end{align*}
and the proof is complete. 
\end{IEEEproof}
Using the next lemma, which relates the value of a 
joint distribution to the values of its marginals, we bound 
$P_\mathrm{code}(v_1^n,v_2^n|u^n)$ in terms of $P(v_1^n,v_2^n|u^n)$. 

\begin{lemma} For every $(u^n,v_1^n,v_2^n)\in A_\delta^{(n)}$, \label{lem:marginal}
\begin{equation*}
  2^{n(I(V_1;V_2|U)-4\delta)}\leq\frac{P(v_1^n,v_2^n|u^n)}{P(v_1^n|u^n)P(v_2^n|u^n)}
  \leq 2^{n(I(V_1;V_2|U)+4\delta)}.
\end{equation*} 
\end{lemma}
\begin{IEEEproof} For every $(u^n,v_1^n,v_2^n)\in A_\delta^{(n)}$, we have
\begin{align*} 
  \frac{P(v_1^n,v_2^n|u^n)}{P(v_1^n|u^n)P(v_2^n|u^n)} 
  &= \frac{P(u^n)P(u^n,v_1^n,v_2^n)}{P(u^n,v_1^n)P(u^n,v_2^n)}\\
  &\leq \frac{2^{-n(H(U)-\delta)}2^{-n(H(U,V_1,V_2)-\delta)}}
  {2^{-n(H(U,V_1)+\delta)}2^{-n(H(U,V_2)+\delta)}}\\
  &= 2^{n(I(V_1;V_2|U)+4\delta)}.
\end{align*}
The lower bound is proved similarly. 
\end{IEEEproof}

\begin{corollary} \label{cor:pbounds} 
For all $(u^n,v_1^n,v_2^n)\in A_\delta^{(n)}$,
\begin{equation*}
  P_\mathrm{code}(v_1^n,v_2^n|u^n)
  \leq 2^{n(\zeta+4\delta)}P(v_1^n,v_2^n|u^n),
\end{equation*}
where $\zeta:=C_{1d}+C_{2d}-I(V_1;V_2|U)$. 
\end{corollary} 

Next we prove upper bounds on the probabilities of 
the error events defined in Section \ref{sec:error}.

\textbf{Bound on }$\pr(\mathcal{E}_0)$: 
From the definition of $\mathcal{E}_0$ 
(Equation (\ref{eq:e0})) it follows
\begin{equation*}
  \pr(\mathcal{E}_0) = 
	1- \sum_{A_\delta^{(n)}}P(u^n)P_\mathrm{code}(v_1^n,v_2^n|u^n).
\end{equation*}	
Let $A_\delta^{(n)}(u^n)$ denote the set of all pairs 
$(v_1^n,v_2^n)$ such that $(u^n,v_1^n,v_2^n)$ is in $A_\delta^{(n)}$.  
Then
\begin{align*}
  \MoveEqLeft
	\sum_{A_\delta^{(n)}}P(u^n)P_\mathrm{code}(v_1^n,v_2^n|u^n)\\
	&= \sum_{A_\delta^{(n)}}P(u^n)
	\sum_{\nu_1,\nu_2}P(\nu_1|u^n)P(\nu_2|u^n)P(v_1^n,v_2^n|u^n,\nu_1,\nu_2)\\
	&= \smashoperator[r]{\sum_{A_\delta^{(n)}(U)}}P(u^n)
	\sum_{\nu_1,\nu_2}\Big(P(\nu_1|u^n)P(\nu_2|u^n)\\
	&\phantom{=\smashoperator[r]{\sum_{A_\delta^{(n)}(U)}}P(u^n)\sum_{\nu_1,\nu_2}\Big(}
	\times\smashoperator{\sum_{A_\delta^{(n)}(u^n)}}P(v_1^n,v_2^n|u^n,\nu_1,\nu_2)
	\Big).\\
\end{align*}
Note that the innermost sum equals
\begin{equation*}
  \sum_{A_\delta^{(n)}(u^n)}P(v_1^n,v_2^n|u^n,\nu_1,\nu_2)
	=\begin{cases}
	1 &\text{if }\mathcal{A}(u^n,\nu_1,\nu_2)\neq\emptyset\\
	0 &\text{otherwise}.
	\end{cases}
\end{equation*}
Thus 
\begin{align*}
  \MoveEqLeft
  \sum_{A_\delta^{(n)}}P(u^n)P_\mathrm{code}(v_1^n,v_2^n|u^n)\\
  &= \sum_{u^n,\nu_1,\nu_2}
	P(u^n)P(\nu_1|u^n)P(\nu_2|u^n)\mathbf{1}\{\mathcal{A}\neq\emptyset\},
\end{align*}
which implies
\begin{align*}
  \pr(\mathcal{E}_0) &= \sum_{u^n,\nu_1,\nu_2}
	P(u^n)P(\nu_1|u^n)P(\nu_2|u^n)\mathbf{1}\{\mathcal{A}=\emptyset\}\\
	&= \pr\big\{\mathcal{A}(U^n,V_1^n(.),V_2^n(.))=\emptyset\big\}.
\end{align*}
The last term goes to zero if $\zeta>4\delta$ (by the Mutual Covering Lemma 
discussed in Appendix \ref{app:mutual}).

\textbf{Bound on }$\pr(\mathcal{E}_1\setminus \mathcal{E}_0)$:
Define the set $B^{(n)}$ as the set
of all $(u^n,v_1^n,v_2^n,x_1^n,x_2^n,y^n)$ where
$(u^n,v_1^n,v_2^n)\in A_\delta^{(n)}$
but 
\begin{equation*}
  (u^n,v_1^n,v_2^n,x_1^n,x_2^n,y^n)\notin 
	A_\epsilon^{(n)}.
\end{equation*}
Then we have 
\begin{align*}
  \pr(\mathcal{E}_1\setminus\mathcal{E}_0)
  &= \sum_{B^{(n)}}
  P(u^n)P_\mathrm{code}(v_1^n,v_2^n|u^n)\\
  &\phantom{=\sum_{\mathcal{B}^{(n)}}}\times
  P(x_1^n|u^n,v_1^n)P(x_2^n|u^n,v_2^n)P(y^n|x_1^n,x_2^n).
\end{align*}
Since the sum is only over all typical triples $(u^n,v_1^n,v_2^n)$,
Corollary \ref{cor:pbounds} implies
\begin{align*}
  \pr(\mathcal{E}_1\setminus\mathcal{E}_0)
  &\leq 2^{n(\zeta+4\delta)}\\
  &\phantom{\leq} \times \sum_{\mathcal{B}^{(n)}}
  \Big(P(u^n)P(v_1^n,v_2^n|u^n)P(x_1^n|u^n,v_1^n)\\
  &\phantom{\leq\times \sum_{\mathcal{B}^{(n)}}\Big(}\times 
  P(x_2^n|u^n,v_2^n)P(y^n|x_1^n,x_2^n)\Big)\\
  &\leq 2^{n(\zeta+4\delta)}
  \pr\big\{(A^{(n)}_\epsilon)^c\big\}\\
	&\leq 2^{n(\zeta+4\delta)}2^{-n\Theta(\epsilon)}.
\end{align*}
Thus 
$\pr(\mathcal{E}_1\setminus\mathcal{E}_0)$ goes to zero if 
$\zeta<\Theta(\epsilon)-4\delta$.
Since 
\begin{equation*}
\pr(\mathcal{E}_0\cup\mathcal{E}_1)=\pr(\mathcal{E}_0)+
\pr(\mathcal{E}_1\setminus\mathcal{E}_0),
\end{equation*} 
$\pr(\mathcal{E}_0\cup\mathcal{E}_1)$
goes to zero if we choose $\delta>0$ and $\zeta>0$ such that 
$\delta<\frac{1}{8}\Theta(\epsilon)$ and $4\delta<\zeta<\Theta(\epsilon)-4\delta$.

\textbf{Bound on }$\pr(\mathcal{E}_U)$: If $\mathcal{E}_U$ occurs, then
\begin{equation*}
  (\hat{U}^n,\hat{V}_1^n,\hat{V}_2^n,\hat{X}_1^n,\hat{X}_2^n,Y^n)
	\in A_\epsilon^{(n)},
\end{equation*}
even though $(\hat{U}^n,\hat{V}_1^n,\hat{V}_2^n,\hat{X}_1^n,\hat{X}_2^n)$
is independent of $(U^n,V_1^n,V_2^n,X_1^n,X_2^n)$ (and thus of $Y^n$)
by our code design. Let $A_\epsilon^{(n)}(Y)$ denote the typical set
with respect to $P(y)$ and suppose $y^n\in A_\epsilon^{(n)}(Y)$. 
Then let $A_\epsilon^{(n)}(y^n)$ denote the set of all 
$(u^n,v_1^n,v_2^n,x_1^n,x_2^n)$ that are jointly typical with $y^n$. 
Then by Theorem 15.2.2 of \cite{CoverThomas}, 
$|A_\epsilon^{(n)}(y^n)|\leq 2^{n(H(U,V_1,V_2,X_1,X_2|Y)+2\epsilon)}$.
Thus $\pr(\mathcal{E}_U)$ is bounded from above by
\begin{equation*}
  2^{n(R_1+R_2)}\sum_{A_\epsilon^{(n)}(Y)}P_\mathrm{code}(y^n)
	\sum_{A_\epsilon^{(n)}(y^n)} P_\mathrm{code}(u^n,v_1^n,v_2^n,x_1^n,x_2^n).
\end{equation*}
We now use Corollary \ref{cor:pbounds} to get
\begin{align*}
  \MoveEqLeft
	\sum_{A_\epsilon^{(n)}(y^n)} P_\mathrm{code}(u^n,v_1^n,v_2^n,x_1^n,x_2^n)\\
	&\leq 2^{n(\zeta+4\epsilon)}\sum_{A_\epsilon^{(n)}(y^n)}P(u^n,v_1^n,v_2^n,x_1^n,x_2^n)\\
	&\leq 2^{n(\zeta+4\epsilon)}\times2^{n(H(U,V_1,V_2,X_1,X_2|Y)+2\epsilon)}\\
	&\phantom{\leq}\times 2^{-n(H(U,V_1,V_2,X_1,X_2)-\epsilon)}\\
	&= 2^{-n(I(X_1,X_2;Y)-\zeta-7\epsilon)}.
\end{align*}
Thus $\pr(\mathcal{E}_U)\rightarrow 0$ if 
\begin{equation*}
  R_1+R_2<I(X_1,X_2;Y)-\zeta-7\epsilon.
\end{equation*}

\textbf{Bound on }$\pr(\mathcal{E}_{V_1X_2})$: If $\mathcal{E}_{V_1X_2}$ occurs, then
\begin{equation*}
  (\hat{w}_{10},\hat{w}_{20},\hat{w}_{2d})
  =(w_{10},w_{20},w_{2d}),
\end{equation*} 
but $\hat{w}_{1d}\neq w_{1d}$ and $\hat{w}_{22}\neq w_{22}$. 
This implies that there are at most $2^{n(R_1-C_{10})^+}$ and 
$2^{n(R_2-C_2^\mathrm{in})^+}$ possible values for $\hat{w}_1$ 
and $\hat{w}_2$, respectively.

In this case
\begin{align*}
  \MoveEqLeft 
  P(v_1^n,v_2^n,\hat{v}_1^n,\hat{v}_2^n|u^n,\nu_2)\\
	&= \sum_{\nu_1,\hat{\nu}_1}\Big(P(\nu_1|u^n)P(\hat{\nu}_1|u^n)\\
	&\phantom{= \sum_{\nu_1,\hat{\nu}_1}\Big(}\times
	P(v_1^n,v_2^n|u^n,\nu_1,\nu_2)P(\hat{v}_1^n,\hat{v}_2^n|u^n,\hat{\nu}_1,\nu_2)\Big)\\
	&= P(v_1^n,v_2^n|u^n,\nu_2)P(\hat{v}_1^n,\hat{v}_2^n|u^n,\nu_2).
\end{align*}
Thus we have the Markov chain
\begin{equation*}
(\hat{V}_1^n,\hat{V}_2^n,\hat{X}_1^n,\hat{X}_2^n)\rightarrow (U^n,V_2^n(.))
\rightarrow Y^n.
\end{equation*}
Therefore we can bound $\pr(\mathcal{E}_{V_1})$ from above by
\begin{align*}
  \MoveEqLeft
  2^{n((R_1-C_{10})^++(R_2-C_2^\mathrm{in})^+)}\\
	&\times
	\sum_{A_\epsilon^{(n)}} \sum_{\nu_2}
	P_\mathrm{code}(u^n,\nu_2,y^n)
	P_\mathrm{code}(v_1^n,v_2^n,x_1^n,x_2^n|u^n,\nu_2),
\end{align*}
We rewrite the sum as 
\begin{align}
  \MoveEqLeft
	\sum_{A_\epsilon^{(n)}} \sum_{\nu_2}
	P_\mathrm{code}(u^n,\nu_2,y^n)
	P_\mathrm{code}(v_1^n,v_2^n,x_1^n,x_2^n|u^n,\nu_2) \label{eq:ev1x2sum}\\
	&= \sum_{A_\epsilon^{(n)}} \Big(P(u^n)P(x_1^n,x_2^n|u^n,v_1^n,v_2^n) \nonumber\\
	&\phantom{= \sum_{A_\epsilon^{(n)}}}\times
	\sum_{\nu_2} P_\mathrm{code}(\nu_2,y^n|u^n)
	P_\mathrm{code}(v_1^n,v_2^n|u^n,\nu_2)\Big). \nonumber
\end{align}
Next, we provide an upper bound for the inner sum. 
\begin{align}
  \MoveEqLeft
	\sum_{\nu_2} P_\mathrm{code}(\nu_2,y^n|u^n)
	P_\mathrm{code}(v_1^n,v_2^n|u^n,\nu_2) \label{eq:ev1x2insum}\\
	&\leq \sum_{\nu_1,\nu_2} P(\nu_1|u^n)
	P_\mathrm{code}(\nu_2,y^n|u^n)
	P_\mathrm{code}(v_1^n,v_2^n|u^n,\nu_1,\nu_2)\nonumber\\
	&\leq \sum_{\nu_1,\nu_2} \Big(P(\nu_1|u^n)P_\mathrm{code}(\nu_2,y^n|u^n)
	\nonumber\\
	&\phantom{\leq \sum_{\nu_1,\nu_2}\Big(} 
	\times \mathbf{1}\big\{\nu_1^{-1}(v_1^n)\neq\emptyset\big\}
	\mathbf{1}\big\{\nu_2^{-1}(v_2^n)\neq\emptyset\big\}\Big),\nonumber
\end{align}
where the last inequality follows from 
\begin{equation*}
  P_\mathrm{code}(v_1^n,v_2^n|u^n,\nu_1,\nu_2)\leq 
	\mathbf{1}\big\{\nu_1^{-1}(v_1^n)\neq\emptyset\big\}
	\mathbf{1}\big\{\nu_2^{-1}(v_2^n)\neq\emptyset\big\}.
\end{equation*}
From the proof of Lemma \ref{lem:pbounds} we have
\begin{equation*}
  \sum_{\nu_1} P(\nu_1|u^n)
	\mathbf{1}\big\{\nu_1^{-1}(v_1^n)\neq\emptyset\big\}
	\leq 2^{nC_{1d}}P(v_1^n|u^n).
\end{equation*}
In addition, we have 
\begin{align*}
  \MoveEqLeft
	\sum_{\nu_2} P_\mathrm{code}(\nu_2,y^n|u^n)
	\mathbf{1}\big\{\nu_2^{-1}(v_2^n)\neq\emptyset\big\}\\
	&= \pr\big\{\exists z:V_2^n(z)=v_2^n,Y^n=y^n|U^n=u^n\big\}\\
	&\leq \sum_{z=1}^{2^{nC_{2d}}}
	\pr\big\{V_2^n(z)=v_2^n,Y^n=y^n|U^n=u^n\big\}\\
	&= 2^{nC_{2d}}P_\mathrm{code}(v_2^n,y^n|u^n),
\end{align*}
where the inequality follows from the union bound. Thus 
\begin{equation*}
  2^{n(C_{1d}+C_{2d})}P(v_1^n|u^n)P_\mathrm{code}(v_2^n,y^n|u^n)
\end{equation*}
is an upper bound for the sum in Equation (\ref{eq:ev1x2insum}). 
We can now bound the sum in Equation (\ref{eq:ev1x2sum}) from above by
\begin{align*}
  \MoveEqLeft
	2^{n(C_{1d}+C_{2d})} \\
	&\times\sum_{A_\epsilon^{(n)}}P_\mathrm{code}(u^n,v_2^n,y^n)
	P(v_1^n|u^n)P(x_1^n,x_2^n|u^n,v_1^n,v_2^n) \\
	&\leq 2^{n(\zeta+4\epsilon)}\sum_{A_\epsilon^{(n)}}P_\mathrm{code}(u^n,v_2^n,y^n)
	P(v_1^n,x_1^n,x_2^n|u^n,v_2^n),
\end{align*}
where the inequality follows from Lemma \ref{lem:marginal}. 

Similar to the notation we used to bound $\pr(\mathcal{E}_0)$
and $\pr(\mathcal{E}_U)$, we define 
$A_\epsilon^{(n)}(U,V_2,Y)$ as the typical set with respect to
$P(u,v_2,y)$. In addition, for every $(u^n,v_2^n,y^n)\in A_\epsilon^{(n)}(U,V_2,Y)$,
we define $A_\epsilon^{(n)}(u^n,v_2^n,y^n)$ as the set of all
$(v_1^n,x_1^n,x_2^n)$ such that 
\begin{equation*}
  (u^n,v_1^n,v_2^n,x_1^n,x_2^n,y^n)\in A_\epsilon^{(n)}.
\end{equation*}
Again by Theorem 15.2.2 of \cite{CoverThomas} we have
\begin{equation*}
  A_\epsilon^{(n)}(u^n,v_2^n,y^n)\leq 
	2^{n(H(V_1,X_1,X_2|U,V_2,Y)+2\epsilon)}.
\end{equation*} 
We can now bound $\pr(\mathcal{E}_{V_1})$ from above by
\begin{align*}
  \MoveEqLeft
  2^{n((R_1-C_{10})^++(R_2-C_2^\mathrm{in})^++\zeta+4\epsilon)}\\
	&\times
	\sum P_\mathrm{code}(u^n,v_2^n,y^n)
	\sum P(v_1^n,x_1^n,x_2^n|u^n,v_2^n),
\end{align*}
where the first sum is over all $(u^n,v_2^n,y^n)$
in $A_\epsilon^{(n)}(U,V_2,Y)$ and the second sum 
is over all $(v_1^n,x_1^n,x_2^n)$ in 
$A_\epsilon^{(n)}(u^n,v_2^n,y^n)$.
We have 
\begin{align*}
  \MoveEqLeft
	\sum_{A_\epsilon^{(n)}(u^n,v_2^n,y^n)}
	P(v_1^n,x_1^n,x_2^n|u^n,v_2^n)\\
	&\leq 2^{n(H(V_1,X_1,X_2|U,V_2,Y)+2\epsilon)}
	2^{-n(H(V_1,X_1,X_2|U,V_2)-2\epsilon)}\\
	&= 2^{-n(I(X_1,X_2;Y|U,V_2)-4\epsilon)}.
\end{align*}
Thus $\pr(\mathcal{E}_{V_1X_2})\rightarrow 0$ if
\begin{equation*}
  (R_1-C_{10})^++(R_2-C_2^\mathrm{in})^+<I(X_1,X_2;Y|U,V_2)-\zeta-8\epsilon.
\end{equation*}

\textbf{Bound on }$\pr(\mathcal{E}_{V_1})$: When $\mathcal{E}_{V_1}$
occurs,
\begin{align*}
  \MoveEqLeft
  \Big(U^n(w_{10},w_{20}),V_1^n(\hat{w}_{1d},\hat{Z}_1),V_2^n(w_{2d},\hat{Z}_2),\\
	&X_1^n(\hat{w}_{11}|U^n,\hat{V}_1^n),X_2^n(w_{22}|U^n,\hat{V}_2^n),
	Y^n\Big)\in A_\epsilon^{(n)}
\end{align*}
for some $\hat{w}_{1d}\neq w_{1d}$. In this case, 
$(\hat{V}_1^n,\hat{V}_2^n,\hat{X}_1^n,\hat{X}_2^n)$ and
$(V_1^n,V_2^n,X_1^n,X_2^n)$ are independent given $(U^n,V_2^n(.),X_2^n(.))$.
Therefore,
\begin{equation*}
  (\hat{V}_1^n,\hat{V}_2^n,\hat{X}_1^n,\hat{X}_2^n)
	\rightarrow (U^n,V_2^n(.),X_2^n(.)) \rightarrow Y^n
\end{equation*} 
is a Markov chain. Thus we can bound $\pr(\mathcal{E}_{V_1})$ from above by
\begin{align*}
  \MoveEqLeft 2^{n(R_1-C_{10})^+}
	\sum_{A_\epsilon^{(n)}}\sum_{\nu_2,\chi_2}
	\Big(P_\mathrm{code}(u^n,\nu_2,\chi_2,y^n)\\
	&\phantom{\sum_{A_\epsilon^{(n)}\sum_{\nu_2,\chi_2}}
	\sum_{\nu_2,\chi_2}\Big(}
	\times P_\mathrm{code}(v_1^n,v_2^n,x_1^n,x_2^n|u^n,\nu_2,\chi_2)\Big)
\end{align*}
We simplify the sum as
\begin{align}
  \MoveEqLeft 
	\sum_{A_\epsilon^{(n)}}\sum_{\nu_2,\chi_2}
	P_\mathrm{code}(u^n,\nu_2,\chi_2,y^n)
	P_\mathrm{code}(v_1^n,v_2^n,x_1^n,x_2^n|u^n,\nu_2,\chi_2) \nonumber\\
	&=\sum_{A_\epsilon^{(n)}}\Big(P(u^n)P(x_1^n|u^n,v_1^n) \label{eq:ev1sum}\\
	&\times
	\sum_{\nu_2,\chi_2} P_\mathrm{code}(\nu_2,\chi_2,y^n|u^n)
	P_\mathrm{code}(v_1^n,v_2^n,x_2^n|u^n,\nu_2,\chi_2)\Big). \nonumber
\end{align}
Next, we find an upper bound on the inner sum. We have 
\begin{align}
  \MoveEqLeft 
	\sum_{\nu_2,\chi_2} P_\mathrm{code}(\nu_2,\chi_2,y^n|u^n)
	P_\mathrm{code}(v_1^n,v_2^n,x_2^n|u^n,\nu_2,\chi_2)\nonumber\\
	&=\sum_{\nu_1,\nu_2,\chi_2} \Big(P_\mathrm{code}(\nu_2,\chi_2,y^n|u^n)
	P(\nu_1|u^n) \nonumber \\
	&\phantom{=\sum_{\nu_1,\nu_2,\chi_2}\Big(}\times 
	P_\mathrm{code}(v_1^n,v_2^n,x_2^n|u^n,\nu_1,\nu_2,\chi_2)\Big)\nonumber \\
	&\leq \sum_{\nu_1,\nu_2,\chi_2} \Big(P_\mathrm{code}(\nu_2,\chi_2,y^n|u^n)
	P(\nu_1|u^n) \label{eq:ev1insum}\\
	&\times 
	\mathbf{1}\big\{\nu_1^{-1}(v_1^n)\neq\emptyset\big\}
	\mathbf{1}\big\{\nu_2^{-1}(v_2^n)\cap\chi_2^{-1}(x_2^n)\neq\emptyset\big\}
	\Big), \nonumber
\end{align}
where the last inequality follows from
\begin{align*}
  \MoveEqLeft P_\mathrm{code}(v_1^n,v_2^n,x_2^n|u^n,\nu_1,\nu_2,\chi_2)\\
	&\leq \mathbf{1}\big\{\nu_1^{-1}(v_1^n)\neq\emptyset\big\}
	\mathbf{1}\big\{\nu_2^{-1}(v_2^n)\cap\chi_2^{-1}(x_2^n)\neq\emptyset\big\}.
\end{align*}
From the proof of Lemma \ref{lem:pbounds}, we get
\begin{equation*}
  \sum_{\nu_1} P(\nu_1|u^n)
	\mathbf{1}\big\{\nu_1^{-1}(v_1^n)\neq\emptyset\big\}
	\leq 2^{nC_{1d}}P(v_1^n|u^n).
\end{equation*}
In addition,  
\begin{align*}
  \MoveEqLeft
	\sum_{\nu_2} P_\mathrm{code}(\nu_2,\chi_2,y^n|u^n)
	\mathbf{1}\big\{\nu_2^{-1}(v_2^n)\cap\chi_2^{-1}(x_2^n)\neq\emptyset\big\}\\
	&= \pr\big\{\exists z:V_2^n(z)=v_2^n,X_2^n(z)=x_2^n,Y^n=y^n|U^n=u^n\big\}\\
	&\leq \smashoperator[l]{\sum_{z=1}^{2^{nC_{2d}}}}
	\pr\big\{V_2^n(z)=v_2^n,X_2^n(z)=x_2^n,Y^n=y^n|U^n=u^n\big\}\\
	&= 2^{nC_{2d}}P_\mathrm{code}(v_2^n,x_2^n,y^n|u^n),
\end{align*}
where the inequality follows from the union bound. Thus 
\begin{equation*}
  2^{n(C_{1d}+C_{2d})}P(v_1^n|u^n)P_\mathrm{code}(v_2^n,x_2^n,y^n|u^n)
\end{equation*}
is an upper bound for the sum in Equation (\ref{eq:ev1insum}). 
We can now bound the sum in Equation (\ref{eq:ev1sum}) from above by
\begin{align*}
  \MoveEqLeft
	2^{n(C_{1d}+C_{2d})} \\
	&\times\sum_{A_\epsilon^{(n)}}P_\mathrm{code}(u^n,v_2^n,x_2^n,y^n)
	P(v_1^n|u^n)P(x_1^n|u^n,v_1^n) \\
	&\leq 2^{n(\zeta+4\epsilon)}\sum_{A_\epsilon^{(n)}}P_\mathrm{code}(u^n,v_2^n,x_2^n,y^n)
	P(v_1^n,x_1^n|u^n,v_2^n),
\end{align*}
where the inequality follows from Lemma \ref{lem:marginal}. 

Finally, we bound $\pr(\mathcal{E}_{V_1})$ from above by
\begin{align*}
  \MoveEqLeft
  2^{n((R_1-C_{10})^++\zeta+4\epsilon)}\\
	&\times
	\sum P_\mathrm{code}(u^n,v_2^n,x_2^n,y^n)
	\sum P(v_1^n|u^n,v_2^n)P(x_1^n|u^n,v_1^n),
\end{align*}
where the first sum is over all $(u^n,v_2^n,x_2^n,y^n)$
in $A_\epsilon^{(n)}(U,V_2,X_2,Y)$ and the second sum 
is over all $(v_1^n,x_1^n)$ in 
$A_\epsilon^{(n)}(u^n,v_2^n,x_2^n,y^n)$.
We have 
\begin{align*}
  \MoveEqLeft
	\sum_{A_\epsilon^{(n)}(u^n,v_2^n,x_2^n,y^n)}
	P(v_1^n,x_1^n|u^n,v_2^n)\\
	&\leq 2^{n(H(V_1,X_1|U,V_2,X_2,Y)+2\epsilon)}
	2^{-n(H(V_1,X_1|U,V_2)-2\epsilon)}\\
	&= 2^{-n(I(X_1;Y|U,V_2,X_2)-4\epsilon)}.
\end{align*}
Thus $\pr(\mathcal{E}_{V_1})\rightarrow 0$ if
\begin{equation*}
  (R_1-C_{10})^+<I(X_1;Y|U,V_2,X_2)-\zeta-8\epsilon.
\end{equation*}

\textbf{Bound on }$\pr(\mathcal{E}_{V_1V_2})$:
The event $\mathcal{E}_{V_1V_2}$ occurs when
$(\hat{w}_{10},\hat{w}_{20})=(w_{10},w_{20})$,
but $\hat{w}_{1d}\neq w_{1d}$ and 
$\hat{w}_{2d}\neq w_{2d}$. In this case, $Y^n$
is independent of 
$(\hat{V}_1^n,\hat{V}_2^n,\hat{X}_1^n,\hat{X}_2^n)$
given $U^n$. This leads to the upper bound
\begin{align*}
  \MoveEqLeft
  2^{n((R_1-C_{10})^++(R_2-C_{20})^+)}\\
	&\times \sum P_\mathrm{code}(u^n,y^n)
	\sum P_\mathrm{code}(v_1^n,v_2^n,x_1^n,x_2^n|u^n)
\end{align*}
for $\pr(\mathcal{E}_{V_1V_2})$, where the sums are over
$A_\epsilon^{(n)}(U,Y)$ and $A_\epsilon^{(n)}(u^n,y^n)$, 
respectively. By Corollary \ref{cor:pbounds}
we have 
\begin{align*}
  \MoveEqLeft
  \sum_{A_\epsilon^{(n)}(u^n,y^n)}
	P_\mathrm{code}(v_1^n,v_2^n,x_1^n,x_2^n|u^n)\\
	&\leq 2^{n(H(V_1,V_2,X_1,X_2|U,Y)+2\epsilon)}\\
  &\phantom{\leq} \times
	2^{-n(-\zeta+H(V_1,V_2,X_1,X_2|U)-6\epsilon)}\\
	&= 2^{-n(I(X_1,X_2;Y|U)-\zeta-8\epsilon)}.
\end{align*}
Hence $\pr(\mathcal{E}_{V_1V_2})$ goes to zero if 
\begin{equation*}
  (R_1-C_{10})^++(R_2-C_{20})^+ < I(X_1,X_2;Y|U)-\zeta-8\epsilon.
\end{equation*}

\textbf{Bound on }$\pr(\mathcal{E}_{X_1})$: If $\mathcal{E}_{X_1}$ occurs,
then
\begin{equation*}
  (\hat{w}_{10},\hat{w}_{20},\hat{w}_{1d},\hat{w}_{2d},\hat{w}_{22})
	=(w_{10},w_{20},w_{1d},w_{2d},w_{22}),
\end{equation*}
but $\hat{w}_{11}\neq w_{11}$ and 
\begin{equation*}
  \big(U^n,V_1^n,V_2^n,\hat{X}_1^n,X_2^n,Y^n\big)\in A_\epsilon^{(n)}.
\end{equation*}
In this case, $\hat{X}_1^n$ and $Y^n$ are independent given
$(U^n,V_1^n,V_2^n,X_2^n)$. Thus we can bound $\pr(\mathcal{E}_{X_1})$ 
from above by 
\begin{equation*}
	2^{n(R_1-C_1^\mathrm{in})^+}\sum
	P_\mathrm{code}(u^n,v_1^n,v_2^n,x_2^n,y^n)
  \sum P(x_1^n|u^n,v_1^n),
\end{equation*}
where the first sum is over $A_\epsilon^{(n)}(U,V_1,V_2,X_2,Y)$ and the
second sum is over $A_\epsilon^{(n)}(u^n,v_1^n,v_2^n,x_2^n,y^n)$.
Further, we have
\begin{equation*}
  \smashoperator[r]{\sum_{A_\epsilon^{(n)}(u^n,v_1^n,v_2^n,x_2^n,y^n)}}
  \quad P(x_1^n|u^n,v_1^n)
	\leq 2^{-n(I(X_1;Y|U,V_1,V_2,X_2)-4\epsilon)},
\end{equation*}
thus $\pr(\mathcal{E}_{X_1})\rightarrow 0$ if
\begin{equation*}
  (R_1-C_1^\mathrm{in})^+ < I(X_1;Y|U,V_1,V_2,X_2)-4\epsilon.
\end{equation*}

\textbf{Bound on }$\pr(\mathcal{E}_{X_1X_2})$: When $\mathcal{E}_{X_1X_2}$
occurs,
\begin{equation*} 
  (\hat{w}_{10},\hat{w}_{20},\hat{w}_{1d},\hat{w}_{2d})
	=(w_{10},w_{20},w_{1d},w_{2d}),
\end{equation*}
but $\hat{w}_{ii}\neq w_{ii}$ for $i=1,2$, and
\begin{equation*}
  \big(U^n,V_1^n,V_2^n,\hat{X}_1^n,\hat{X}_2^n,Y^n\big)\in A_\epsilon^{(n)}.
\end{equation*}
In this case $Y^n$ is independent of $(\hat{X}_1^n,\hat{X}_2^n)$ given
$(U^n,V_1^n,V_2^n)$. Thus $\pr(\mathcal{E}_{X_1X_2})$ is bounded by
\begin{align*}
  \MoveEqLeft
	2^{n((R_1-C_1^\mathrm{in})^++(R_2-C_2^\mathrm{in})^+)}\\
	&\times
	\sum P_\mathrm{code}(u^n,v_1^n,v_2^n,y^n)
	\sum P(x_1^n,x_2^n|u^n,v_1^n,v_2^n)
\end{align*}
where the first sum is over $A_\epsilon^{(n)}(U,V_1,V_2,Y)$ 
and the second sum is over $A_\epsilon^{(n)}(u^n,v_1^n,v_2^n,y^n)$.
We have 
\begin{align*}
  \smashoperator[r]{\sum_{A_\epsilon^{(n)}(u^n,v_1^n,v_2^n,y^n)}} 
	P(x_1^n,x_2^n|u^n,v_1^n,v_2^n)
	\leq 2^{-n(I(X_1,X_2;Y|U,V_1,V_2)-4\epsilon)},
\end{align*}
thus $\pr(\mathcal{E}_{X_1X_2})$ goes to zero if 
\begin{equation*}
  (R_1-C_1^\mathrm{in})^++(R_2-C_2^\mathrm{in})^+
	< I(X_1,X_2;Y|U,V_1,V_2)-4\epsilon.
\end{equation*}

\section{The Mutual Covering Lemma} \label{app:mutual}
In this appendix, we state and prove the mutual covering lemma, 
which is a variation of a result by the same name in
the book by El Gamal and Kim \cite{ElGamalKim}.
Our result differs from the result in \cite{ElGamalKim} in two ways. One, our result 
is stated and proven for weakly typical sets, rather than strongly typical 
sets, and two, we require complete independence rather than
pairwise independence between codewords. 

\begin{lemma} [Mutual Covering Lemma] \label{lem:covering} Let $U$, $V_{1}$, and $V_{2}$ 
be random variables jointly distributed
as $P(u,v_1,v_2)$. Suppose $\mathcal{A}$ and $\mathcal{B}$ are finite sets 
with $|\mathcal{A}|\geq 2^{nr_1}$ and $|\mathcal{B}|\geq 2^{nr_2}$.
Given $U^n=u^n$, for every $(a,b)\in\mathcal{A}\times\mathcal{B}$, 
let $V^n_1(a)$ and $V^n_2(b)$ be random vectors 
generated in an i.i.d.\ manner according to the distributions
\begin{align*}
  \pr\big\{V_1^n(a)=v_1^n|U^n=u^n\big\} &= \prod_{t=1}^n P(v_{1t}|u_t)\\
  \pr\big\{V_2^n(b)=v_2^n|U^n=u^n\big\} &= \prod_{t=1}^n P(v_{2t}|u_t),
\end{align*}
where $P(v_1|u)$ and $P(v_2|u)$ are the marginals of $P(v_1,v_2|u)$.
Then
\begin{equation*}
  \lim_{n\rightarrow\infty}\pr
  \Big\{\exists (a,b)\in\mathcal{A}\times\mathcal{B}:
  (U^n,V_1^n(a),V_2^n(b))\in A_\delta^{(n)}\Big\}=1
\end{equation*}
if $r_1+r_2>I(V_{1};V_{2}|U)+4\delta$.
\end{lemma}

Our proof, which is given in detail at the end of this appendix,  
follows the achievability proof of the rate-distortion theorem
given in \cite[pp. 318-324]{CoverThomas}. 

The next corollary follows from Lemma \ref{lem:marginal} 
in Appendix \ref{app:error} and is the conditional version of Lemma 10.5.2
of \cite{CoverThomas}.
\begin{corollary} \label{cor:marginal}
For every $(u^n,v_1^n,v_2^n)\in A_\delta^{(n)}$,
\begin{equation*}
  P(v_2^n|u^n)\geq P(v_2^n|u^n,v_1^n)2^{-n(I(V_1;V_2|U)+4\delta)}.
\end{equation*}
\end{corollary}
We next prove the Mutual Covering Lemma. It suffices to show
\begin{equation*}
  \lim_{n\rightarrow\infty}
	\pr\Big\{\forall (a,b)\in\mathcal{A}\times\mathcal{B}:
	(U^n,V_1^n(a),V_2^n(b))\notin A_\delta^{(n)}\Big\}
	=0.
\end{equation*}
For every $(u^n,v_1^n,v_2^n)$, define
\begin{equation*}
  K(u^n,v_1^n,v_2^n)=\begin{cases}
    1 & \text{if }(u^n,v_1^n,v_2^n)\in A_{\delta}^{(n)},\\
    0 & \text{otherwise}. 
  \end{cases}
\end{equation*}
Then we have
\begin{align}
  \MoveEqLeft \pr\Big\{\forall (a,b)\in\mathcal{A}\times\mathcal{B}:
  (U^n,V_1^n(a),V_2^n(b))\notin A_\delta^{(n)}\Big\} \nonumber \\	
  &= \sum_{u^n}P(u^n)\Big[1-\sum_{v_1^n,v_2^n}
  \Big(K(u^n,v_1^n,v_2^n) \nonumber \\
  &\phantom{\leq \sum_{u^n}P(u^n)\Big[1-\sum_{v_1^n,v_2^n}
  \Big(}\times
  P(v_1^n|u^n)P(v_2^n|u^n)\Big)\Big]^{|\mathcal{A}||\mathcal{B}|} \nonumber \\
  &\leq \sum_{u^n}P(u^n)\Big[1-2^{-n(I(V_1;V_2|U)+4\delta)}\sum_{v_1^n,v_2^n}
  \Big(K(u^n,v_1^n,v_2^n) \nonumber \\
  &\phantom{\leq \sum_{u^n}P(u^n)\Big[}\times
  P(v_1^n|u^n)P(v_2^n|u^n,v_1^n)\Big)\Big]^{|\mathcal{A}||\mathcal{B}|} \label{eq:a}
\end{align}
where the inequality follows by Corollary \ref{cor:marginal}. 
By Lemma 10.5.3 of \cite{CoverThomas}, which states that for
$x,y\in [0,1]$ and positive $n$,
\begin{equation*}
  (1-xy)^n\leq 1-x+e^{-yn},
\end{equation*}
the right
hand side of Equation (\ref{eq:a}) can be bounded from above by
\begin{align*}
  \MoveEqLeft
  \sum_{u^n}P(u^n)\Big[1-\sum_{v_1^n,v_2^n}K(u^n,v_1^n,v_2^n)
  P(v_1^n,v_2^n|u^n)\\
  &\phantom{\sum_{u^n}P\Big[}
  +\exp\big(-|\mathcal{A}||\mathcal{B}|2^{-n(I(V_1;V_2|U)+4\delta)}\big)\Big]\\	
  &\leq 1-\sum_{u^n,v_1^n,v_2^n}K(u^n,v_1^n,v_2^n)P(u^n,v_1^n,v_2^n)\\
  &\phantom{\leq}+\exp\big(-2^{n(r_1+r_2-I(V_1;V_2|U)-4\delta)}\big)\\
	&= \pr\big\{(U^n,V_1^n,V_2^n)\notin A_\delta^{(n)}\big\}\\
  &\phantom{=}+\exp\big(-2^{n(r_1+r_2-I(V_1;V_2|U)-4\delta)}\big)
\end{align*}
which goes to zero as $n\rightarrow \infty$ provided 
$r_1+r_2>I(V_1;V_2|U)+4\delta$. 

\section{Proof of Theorem \ref{thm:gaussian}}\label{app:gaussian}
Note that if $(S,T)$ is a jointly Gaussian random pair
with correlation matrix 
\begin{equation*}
  \Sigma_{ST}=\begin{pmatrix}
  \sigma^2_S & \rho\sigma_S\sigma_T \\
  \rho\sigma_S\sigma_T  & \sigma^2_T
 \end{pmatrix},
\end{equation*}
then by Theorem 8.4.1 of \cite{CoverThomas},
\begin{align} 
  H(S,T) &= \frac{1}{2}\log\big|2\pi e\Sigma_{ST}\big| \nonumber\\
	&=\frac{1}{2}\log \Big((2\pi e)^2(1-\rho^2)\sigma^2_S\sigma^2_T\Big) \label{eq:entropy}
\end{align} 
and 
\begin{align} 
  H(S|T) &= H(S,T)-H(T) \nonumber\\
	&=\frac{1}{2}\log \Big((2\pi e)(1-\rho^2)\sigma^2_S\Big). \label{eq:centropy}
\end{align} 
Choose $(C_{10},C_{20})$ and 
$(\rho_0,\rho_{10},\rho_{20},\rho_{1d},\rho_{2d})$
such that the constraints of Theorem \ref{thm:gaussian} are satisfied.
Then choose $U$, $(V_1,V_2)$, and $(X'_1,X'_2)$ independently according
to the distributions $U\sim\mathcal{N}(0,1)$, $(V_1,V_2)\sim\mathcal{N}(\mathbf{0},\Sigma)$,
and $(X'_1,X'_2)\sim\mathcal{N}(\mathbf{0},\mathrm{I}_2)$, where
\begin{equation*}
  \Sigma=\begin{pmatrix}
  1 & \rho_0 \\
  \rho_0 & 1
 \end{pmatrix}
\end{equation*}
and $\mathrm{I}_2$ is the $2\times 2$ identity matrix. Then
\begin{equation*}
  I(V_1;V_2|U) = I(V_1;V_2)
	= \frac{1}{2}\log\frac{1}{1-\rho_0^2}.
\end{equation*}
Next, for $i=1,2$, define
\begin{equation*}
  \frac{1}{\sqrt{P_i}}X_i=\rho_{i0}U+\rho_{id}V_i+\rho_{ii}X'_i.
\end{equation*}
Note that by Equation (\ref{eq:rii}), this definition results in $\mathbb{E}[X_i^2]=P_i$
for $i=1,2$. Since $Y=X_1+X_2+Z$, 
\begin{align*}
  Y &= (\rho_{10}\sqrt{P_1}+\rho_{20}\sqrt{P_2})U
	+\rho_{1d}\sqrt{P_1}V_1+\rho_{2d}\sqrt{P_2}V_2\\
	&\phantom{\leq}+\rho_{11}\sqrt{P_1}X'_1+\rho_{22}\sqrt{P_2}X'_2+Z.
\end{align*}
Next, we use Equations (\ref{eq:entropy}) and (\ref{eq:centropy})
to calculate the bounds in Theorem \ref{thm:main}, to obtain 
Theorem \ref{thm:gaussian}. In the calculations that follow
let $\{i,j\}=\{1,2\}$. We have
\begin{align*}
  I(X_i;Y|U,V_1,V_2,X_j)
	&= H(Y|U,V_1,V_2,X_j)-H(Y|X_1,X_2)\\
	&= H(\rho_{ii}\sqrt{P_i}X'_i+Z)-H(Z)\\
	&= \frac{1}{2}\log (1+\rho_{ii}^2\gamma_i).
\end{align*}
Next we calculate
\begin{align*}
  \MoveEqLeft I(X_1,X_2;Y|U,V_1,V_2)\\
	&= H(Y|U,V_1,V_2)-H(Y|X_1,X_2)\\
	&= H(\rho_{11}\sqrt{P_1}X'_1+\rho_{22}\sqrt{P_2}X'_2+Z)-H(Z)\\
	&= \frac{1}{2}\log(1+\rho_{11}^2\gamma_1+\rho_{22}^2\gamma_2).
\end{align*}

Unlike the above calculations, for which we only required Equation
(\ref{eq:entropy}), in the calculation of the next two mutual information terms
we require Equation (\ref{eq:centropy}), since the random variables which appear
in the corresponding conditional entropies are dependent. We have
\begin{align*}
  \MoveEqLeft I(X_i;Y|U,V_j,X_j)\\
	&= H(Y|U,V_j,X_j)-H(Y|X_1,X_2)\\
	&= H\big(\rho_{id}\sqrt{P_i}V_i+\rho_{ii}\sqrt{P_i}X'_i+Z|V_j\big)
	-H(Z)\\
	&= \frac{1}{2}\log \big(1+(\rho_{id}^2+\rho_{ii}^2)\gamma_i\big)
	\Big(1-\frac{\rho_0^2\rho_{id}^2\gamma_1}{1+(\rho_{id}^2+\rho_{ii}^2)\gamma_i}\Big)\\
	&= \frac{1}{2} \log(1+\tilde{\rho}_{ii}^2\gamma_i)
\end{align*}
and
\begin{align*}
  \MoveEqLeft I(X_1,X_2;Y|U,V_i)\\
	&= H(Y|U,V_i)-H(Y|X_1,X_2)\\
	&= H\big(\rho_{jd}\sqrt{P_j}V_j+
	\rho_{11}\sqrt{P_1}X'_1+\rho_{22}
	\sqrt{P_2}X'_2+Z|V_i\big)\\
	&\phantom{=}-H(Z)\\
	&=\frac{1}{2}\log\bigg[\big(1+\rho_{ii}^2\gamma_i+(1-\rho_{j0}^2)
	\gamma_j\big)\\
	&\phantom{=\frac{1}{2}\log\bigg[}\times\Big(1-\frac{\rho_0^2\rho_{jd}^2\gamma_j}
	{1+\rho_{ii}^2\gamma_i+(1-\rho_{j0}^2)\gamma_j}\Big)\bigg]\\
	&= \frac{1}{2}\log(1+\rho_{ii}^2\gamma_i+\tilde{\rho}_{jj}^2\gamma_j).
\end{align*}
For the final two remaining expressions, we have
\begin{align*}
  I(X_1,X_2;Y|U) &= H(Y|U)-H(Z)\\
	&= \frac{1}{2}\log \big(1+(1-\rho_{10}^2)\gamma_1
	+(1-\rho_{20}^2)\gamma_2\\
	&\phantom{=\frac{1}{2}\log \big(} 
	+2\rho_0\rho_{1d}\rho_{2d}\bar{\gamma}\big)
\end{align*}
and 
\begin{align*}
  I(X_1,X_2;Y) &= H(Y)-H(Z)\\
	&= \frac{1}{2}\log \big(1+\gamma_1
	+\gamma_2	+2(\rho_{10}\rho_{20}+\rho_0\rho_{1d}\rho_{2d})
	\bar{\gamma}\big),
\end{align*}
where $\bar{\gamma}=\sqrt{\gamma_1\gamma_2}$.

\bibliographystyle{IEEEtran}
\bibliography{ref}

\end{document}